\documentclass[%
 reprint,
 superscriptaddress,
 amsmath,amssymb,
 aps,
 prb,
]{revtex4-2}

\usepackage{graphicx}
\usepackage{subfigure}
\usepackage{float}
\usepackage{dcolumn}
\usepackage{bm}
\usepackage{rotating}
\usepackage{color}
\usepackage{hyperref}

\begin{document}

\preprint{APS/123-QED}

\title{Exact Mobility Edges in a Disorder-Free Dimerized Stark Lattice with Effective Unbounded Hopping}

\author{Yunyao Qi}
    \affiliation{State Key Laboratory of Low-Dimensional Quantum Physics and Department of Physics, Tsinghua University, Beijing 100084, China}
\author{Heng Lin}
    \affiliation{State Key Laboratory of Low-Dimensional Quantum Physics and Department of Physics, Tsinghua University, Beijing 100084, China}
\author{Quanfeng Lu}
    \affiliation{State Key Laboratory of Low-Dimensional Quantum Physics and Department of Physics, Tsinghua University, Beijing 100084, China}
\author{Dong Ruan}%
    \email{dongruan@tsinghua.edu.cn}
    \affiliation{State Key Laboratory of Low-Dimensional Quantum Physics and Department of Physics, Tsinghua University, Beijing 100084, China}
    \affiliation{Frontier Science Center for Quantum Information, Beijing 100084, China}
\author{Gui-Lu Long}%
    \email{gllong@tsinghua.edu.cn}
    \affiliation{State Key Laboratory of Low-Dimensional Quantum Physics and Department of Physics, Tsinghua University, Beijing 100084, China}
    \affiliation{Beijing Academy of Quantum Information Sciences, Beijing 100193, China}
    \affiliation{Frontier Science Center for Quantum Information, Beijing 100084, China}
    \affiliation{Beijing National Research Center for Information Science and Technology, Beijing 100084, China}





\begin{abstract}
We propose a disorder-free one-dimensional single-particle Hamiltonian hosting an exact mobility edge (ME), placing the system outside the assumptions of no-go theorems regarding unbounded potentials. By applying a linear Stark potential selectively to one sublattice of a dimerized chain, we generate an effective Hamiltonian with unbounded, staggered hopping amplitudes. The unbounded nature of the hopping places the model outside the scope of the Simon-Spencer theorem, while the staggered scaling allows it to evade broader constraints on Jacobi matrices. We analytically derive the bulk spectrum in reciprocal space, identifying a sharp ME where the energy magnitude equals the intercell hopping strength. This edge separates a continuum of extended states from two distinct localized branches: a standard unbounded Wannier--Stark ladder and an anomalous bounded branch accumulating at the ME. The existence of extended states is supported by finite-size scaling of the inverse participation ratio up to system sizes $L \sim 10^9$. Furthermore, we propose an experimental realization using photonic frequency synthetic dimensions. Our numerical results indicate that the ME is robust against potential experimental imperfections, including frequency detuning errors and photon loss, establishing a practical path for observing MEs in disorder-free systems.
\end{abstract}

\maketitle
\section{\label{sec:Intro} Introduction}
Anderson localization (AL)~\cite{Anderson_1958, Lee_1985, Evers_2008}, the fundamental phenomenon where eigenstates exponentially localize due to disorder, has been a central focus in condensed matter physics for over half a century. AL is governed by the scaling theory~\cite{Abrahams_1979, Wegner_1976, Thouless_1974, Lee_1981}, which dictates that all states in one- and two-dimensional (1D and 2D) systems localize in the presence of arbitrary uncorrelated disorder. This has traditionally restricted the study of mobility edges (MEs)—critical energies separating extended from localized states—to three dimensions~\cite{Mott_1987, Semeghini_2015, Delande_2017, Hannaford_2022, Matis_2022}. To circumvent dimensional constraints, research has pivoted to quasiperiodic 1D systems, such as the Aubry-Andr\'{e}-Harper (AAH) model~\cite{Aubry_1980, Harper_1955}. While the standard AAH model exhibits a global localization transition due to its self-duality, generalized quasiperiodic models breaking this duality can host exact MEs~\cite{Das_1988, Hiramoto_1989, Das_1990, Boers_2007, Biddle_2010, Bodyfelt_2014, Ganeshan_2015, Rossignolo_2019, Deng_2019,Yao_2019, Liu_2020, Wang_2020, Zeng_2021, Zhao_2023, Roy_2021, Wang_2021, RoyShilpi_2021, Duthie_2021, Miguel_2022, Wang_2023two, Vu_2023, Chang_2025, Chang_2025_dimer}. This phenomenon has been confirmed experimentally across multiple platforms, including semiconductor superlattices~\cite{Merlin_1985}, photonics~\cite{Lahini_2009}, ultracold atoms~\cite{Fallani_2007, Roati_2008, Derrico_2014, Michael_2015, Bordia_2016, Bordia_2017, Luschen_2017, An_2018}, cavities~\cite{Tanese_2014, Goblot_2020}, and superconducting circuits~\cite{Roushan_2017, Li_2023}. Novel MEs, which separate localized and extended states from critical states, have also attracted significant interest recently~\cite{Li_2017, Tong_2022, Zhang_2022, Fraxanet_2022, Zhou_2023}.

Among different approaches to realize MEs in low-dimensional systems, we highlight the ``mosaic lattice''~\cite{Wang_2020, Zeng_2021, Zhao_2023, Zhang_2025}, where quasiperiodic modulations apply only to equally spaced sites. Recently, the mosaic concept was applied to realize MEs in \textit{disorder-free} systems (i.e., lacking randomness or quasiperiodicity) subject to a linear Stark potential, known as the mosaic Stark lattice~\cite{MosaicStark_2022}. Rooted in the physics of Wannier--Stark (WS) localization~\cite{Wannier_1962, Fukuyama_1973, Emin_1987, Martin_1996, Hartmann_2004}, this proposal has attracted broad interest~\cite{Zeng_2023, Gao_2023, Qi_2023, Wei_2024, jiang_2024, Zhao_2025}. However, this approach has been challenged by Ref.~\cite{Longhi_2023}, which invoked the rigorous Simon-Spencer theorem~\cite{SimonSpencer_1989} to argue that any ME observed in systems with unbounded potentials is a transient ``pseudo-ME'' that vanishes in the thermodynamic limit. Furthermore, the hyperexponential nature of WS localization renders standard characterization tools like the Lyapunov exponent unreliable~\cite{Longhi_2023}. Consequently, the realization of exact MEs in 1D disorder-free systems remains an open question.

In this paper, we address this question by proposing a mechanism that fundamentally evades the Simon-Spencer prohibition. We introduce a dimerized lattice equivalent to the Su--Schrieffer--Heeger (SSH) model~\cite{SSH_1979}, subject to a linear Stark potential acting exclusively on the $B$ sublattice. We demonstrate that the interplay between lattice chirality and the sublattice-dependent field generates an effective Hamiltonian with \textit{unbounded, staggered} off-diagonal couplings. Specifically, the effective hopping scales asymptotically linearly with the lattice index. This nonuniform and unbounded hopping places the model outside the definition of discrete Schrödinger operators assumed by the Simon-Spencer theorem~\cite{SimonSpencer_1989}. Furthermore, the staggered structure of the hopping allows the system to evade broader conditions prohibiting the absolutely continuous (AC) spectrum in unbounded Jacobi matrices~\cite{Cojuhari_2008, Swiderski_2016}, allowing for the coexistence of extended and localized states.

We analytically derive the bulk spectrum in momentum space, identifying a sharp ME at $|E| = t_2$ (where 
$t_2$ is the intercell hopping in the effective SSH model). Below the ME ($|E| < t_2$), we demonstrate the existence of a continuum of extended states, supported by numerical scaling analysis of the inverse participation ratio (IPR) up to system sizes $L\sim10^9$ (fractal dimension $D_2 \approx 1.00$). Above the ME, the spectrum splits into two localized branches: a standard unbounded WS ladder and a bounded anomalous branch accumulating at the ME.

Finally, we outline an experimental proposal using \textit{photonic frequency synthetic dimensions}~\cite{SyntheticDim_2018}. We show that our model corresponds to a Creutz ladder~\cite{Creutz_1999} with an additional Stark potential, which can be implemented using coupled thin-film lithium niobate (TFLN) ring resonators. By adopting the architecture in Ref.~\cite{PhotonicCreutz_2025} to realize the Creutz ladder, and utilizing free spectral range (FSR) detuning to generate the Stark potential~\cite{BlochOscillation_2016}, we show via simulation that the ME is observable under realistic experimental conditions. We perform a stability analysis against potential imperfections, specifically FSR mismatch and photon loss, indicating that the extended phase remains robust within the bandwidth of current photonic devices.

The paper is organized as follows. Section~II introduces the model and the mapping to the unbounded Jacobi matrix, deriving the bulk spectrum. Section~III presents numerical results validating the phase diagram and large-scale finite-size scaling. Section~IV details the experimental proposal based on the photonic frequency synthetic dimensions and analyzes the system's stability. Finally, we conclude in Sec.~V.

\section{Model and Exact Solution}

\begin{figure}[htbp]
    \centering
    \includegraphics[width=1.0\linewidth]{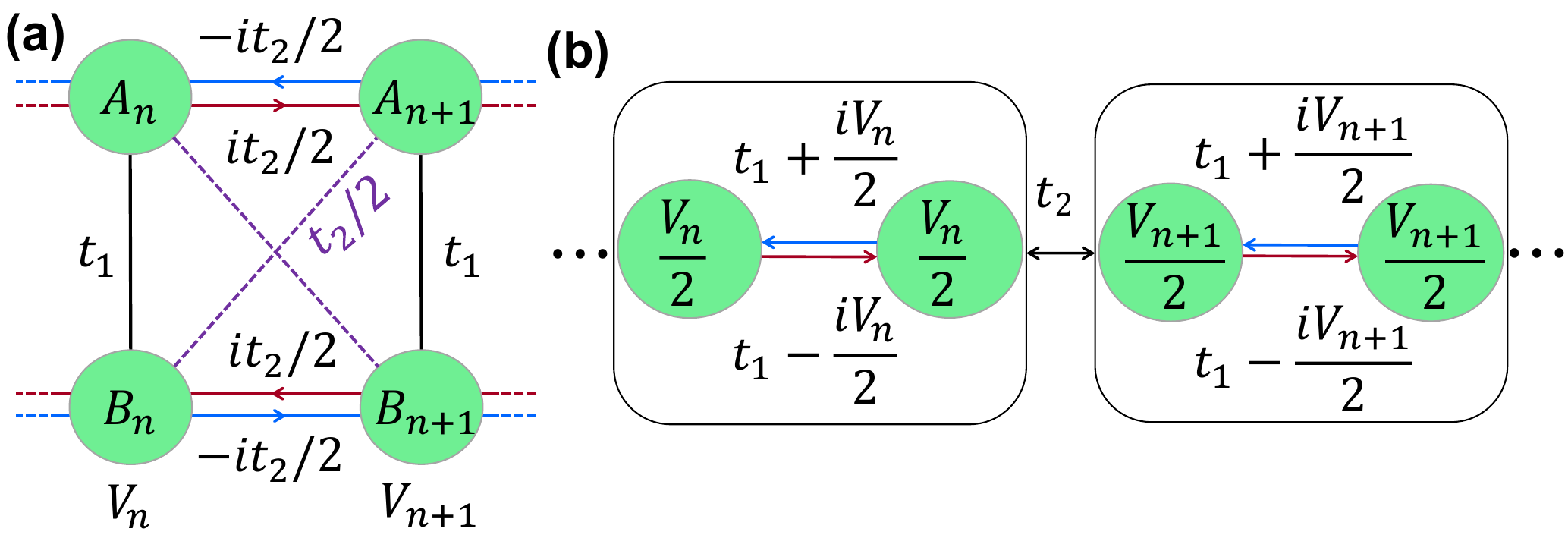}
    
    \par\vspace{0.2cm} 
    
    \begin{minipage}[b]{0.49\linewidth}
        \centering
        \includegraphics[width=\linewidth]{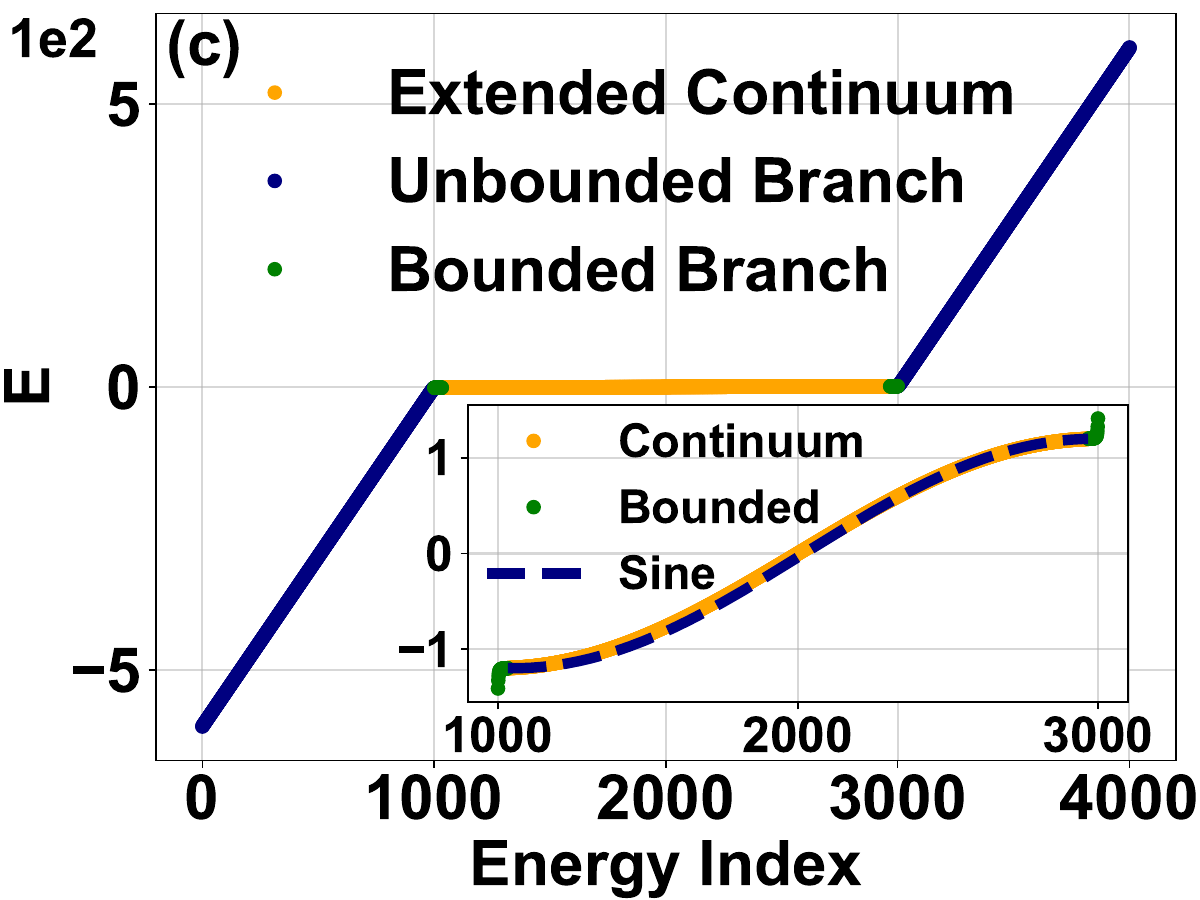}
    \end{minipage}
    \hfill 
    \begin{minipage}[b]{0.49\linewidth}
        \centering
        \includegraphics[width=\linewidth]{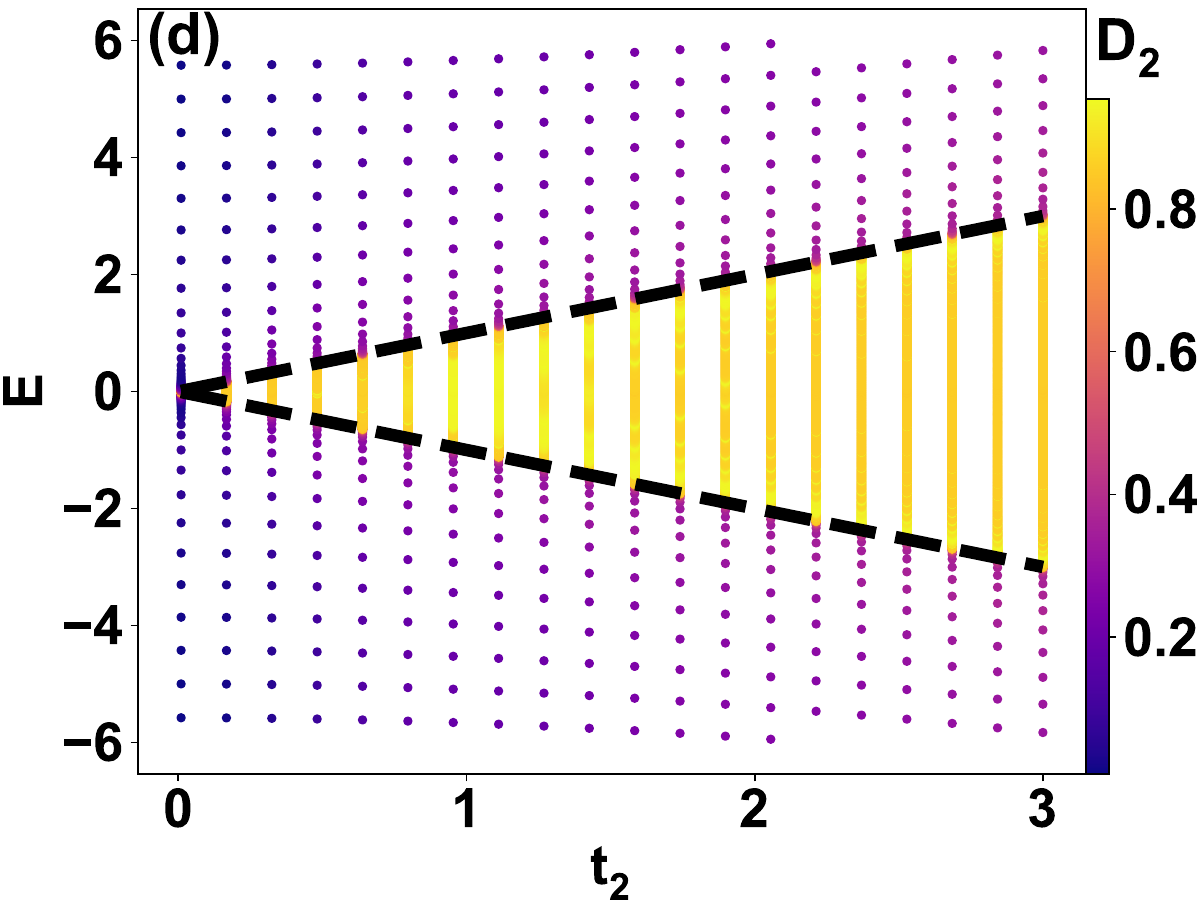}
    \end{minipage}
    \caption{(a) Schematic of the original lattice with linear Stark potential $V_n = -F n$ on the $B$ sublattice. (b) Effective nearest-neighbor hopping lattice after unitary rotation. (c) Sorted energy spectrum for $L=2001$ ($N=1000$), $t_1=1$, $t_2=1.2$, and $F=0.6$ under open boundary conditions. Colors denote the extended continuum (orange), unbounded Wannier--Stark ladder (blue), and bounded anomalous branch (green). Inset: Low-energy zoom and the theoretical dispersion $E \approx t_2 \sin k$ (dashed). (d) Fractal dimension ($D_2$) phase diagram in the ($t_2, E$) plane (truncated at $|E| < 6 t_1$). Dashed lines indicate the analytical mobility edge $|E|=t_2$. Other parameters match those in (c).
}
    \label{fig:main_results}
\end{figure}

We consider a single-particle Hamiltonian on a quasi-1D Creutz ladder [as shown in Fig.~\ref{fig:main_results}(a)] with $L=2N+1$ unit cells, indexed by $n \in \{-N, \dots, N\}$. Each unit cell contains two sublattices, $A$ and $B$. The real-space Schrödinger equations governing the system are given by
\begin{equation}
\begin{aligned}
E\psi_n^A =& t_1\psi_n^B + \frac{t_2}{2} (\psi_{n+1}^B + \psi_{n-1}^B) - \frac{it_2}{2} (\psi_{n+1}^A - \psi_{n-1}^A), \\
E\psi_n^B =& t_1\psi_n^A + \frac{t_2}{2} (\psi_{n+1}^A + \psi_{n-1}^A) + \frac{it_2}{2} (\psi_{n+1}^B - \psi_{n-1}^B) \\
&- F n \psi_n^B.
\end{aligned}
\label{eq:model}
\end{equation}
Here, $t_1$ and $t_2$ represent the intracell and intercell couplings in the equivalent SSH model, respectively, and the term $-F n$ denotes a linear Stark potential applied exclusively to the $B$ sites.

To reveal the underlying structure of the model, we perform a local unitary rotation in the sublattice spinor space, $U = \bigoplus_n e^{-i (\pi/4) \sigma_x}$. This transformation maps the Pauli matrices as $U^\dagger\sigma_z U = \sigma_y$, converting the diagonal Stark potential into a term involving imaginary hopping. Defining the transformed wavefunction $\tilde{\psi} = U^\dagger \psi$, the coupled equations simplify to a nearest-neighbor (NN) form [as shown in Fig.~\ref{fig:main_results}(b)]
\begin{equation}
\begin{aligned}
\left(E + \frac{F n}{2}\right) \tilde{\psi}_n^A &= \left(t_1 - i\frac{F n}{2}\right) \tilde{\psi}_n^B + t_2 \tilde{\psi}_{n-1}^B, \\
\left(E + \frac{F n}{2}\right) \tilde{\psi}_n^B &= \left(t_1 + i\frac{F n}{2}\right) \tilde{\psi}_n^A + t_2 \tilde{\psi}_{n+1}^A.
\end{aligned}
\label{eq:rotated_model}
\end{equation}
The effective model Eq.~(\ref{eq:rotated_model}) highlights two key physical properties. First, the spectrum is symmetric about zero. The effective Hamiltonian maps to its negative under the combined operation of chiral symmetry $\mathcal{S}=\bigoplus_n \sigma_z$ and spatial inversion $\mathcal{P}$ ($n \to -n$, which implies $V_n \to -V_n$). This symmetry $\mathcal{S}\mathcal{P}$ guarantees that the energy eigenvalues appear in pairs $(E, -E)$.

Second, the effective intracell hopping amplitude is complex and site dependent, $J_{\text{intra}}^{\pm}(n) = t_1 \pm iF n/2$. Its magnitude grows asymptotically linearly, matching the potential $F n / 2$. Physically, this growing hopping amplitude provides an effectively expanding bandwidth as the particle moves along the lattice. This enhanced kinetic energy competes directly with the strong localizing tendency of the unbounded Stark potential, ultimately allowing extended states to survive. Meanwhile, the Simon-Spencer theorem regarding unbounded potentials (Ref.~\cite{SimonSpencer_1989}, Theorem 2.1) strictly assumes the hopping terms to be NN
and uniform. While the original Hamiltonian Eq.~(\ref{eq:model}) already falls outside the scope of this theorem due to the presence of next-nearest-neighbor couplings (terms connecting same-sublattice sites $A_n$ and $A_{n\pm 1}$), the effective model explicitly reveals the evasion mechanism. The transformation maps the system to a NN Jacobi matrix where the $n$-dependent, unbounded nature of the hopping violates the uniform assumption of the theorem. Moreover, the staggered nature of the hopping—alternating between unbounded intracell terms and bounded intercell terms ($t_2$)—is crucial for evading more general spectral constraints on unbounded Jacobi matrices~\cite{Cojuhari_2008, Swiderski_2016}. Consequently, our model is mathematically permitted to host a ME in the thermodynamic limit (see Appendix~\ref{app:gauge_transformation} and Appendix~\ref{app:Simon_Spencer} for detailed discussions).

To obtain the bulk spectrum in the thermodynamic limit, we transition to reciprocal space using the Fourier transform. We start from the original Hamiltonian Eq.~\ref{eq:model} without transformation. In this representation, the position operator transforms as $n \to i \partial_k$~\cite{Hartmann_2004}. The Schrödinger equation becomes a system of coupled first-order differential equations
\begin{equation}\label{eq:k_space}
\begin{aligned}
E \phi_A(k) &= t_2 \sin k \, \phi_A(k) + (t_1 + t_2 \cos k) \phi_B(k) , \\
E \phi_B(k) &= (t_1 + t_2 \cos k) \phi_A(k) - (t_2 \sin k + iF \partial_k) \phi_B(k) .
\end{aligned}
\end{equation}
By substituting $\phi_A(k)$ from the first equation into the second, we decouple the system to obtain a single first-order ordinary differential equation for the $B$-sublattice component
\begin{equation}\label{eq:ODE}
\begin{aligned}
iF \frac{d \phi_B}{dk} &= \mathcal{V}(k, E) \phi_B(k), \quad \text{with}\\
\mathcal{V}(k, E) &= \frac{t_1^2 + t_2^2 + 2t_1t_2 \cos k - E^2}{E - t_2 \sin k}.
\end{aligned}
\end{equation}
The spectral properties of the system are determined by the singularity structure of the potential $\mathcal{V}(k, E)$, specifically, the denominator $D(k) = E - t_2 \sin k$. The condition for the appearance of singularities, $|E|\le t_2$, defines the exact ME, which corresponds to the V-shaped phase boundary observed in the numerical phase diagram in Fig.~\ref{fig:main_results}(d).

\begin{figure}[htbp]
    \centering
    \begin{minipage}[b]{0.49\linewidth}
        \centering
        \includegraphics[width=\linewidth]{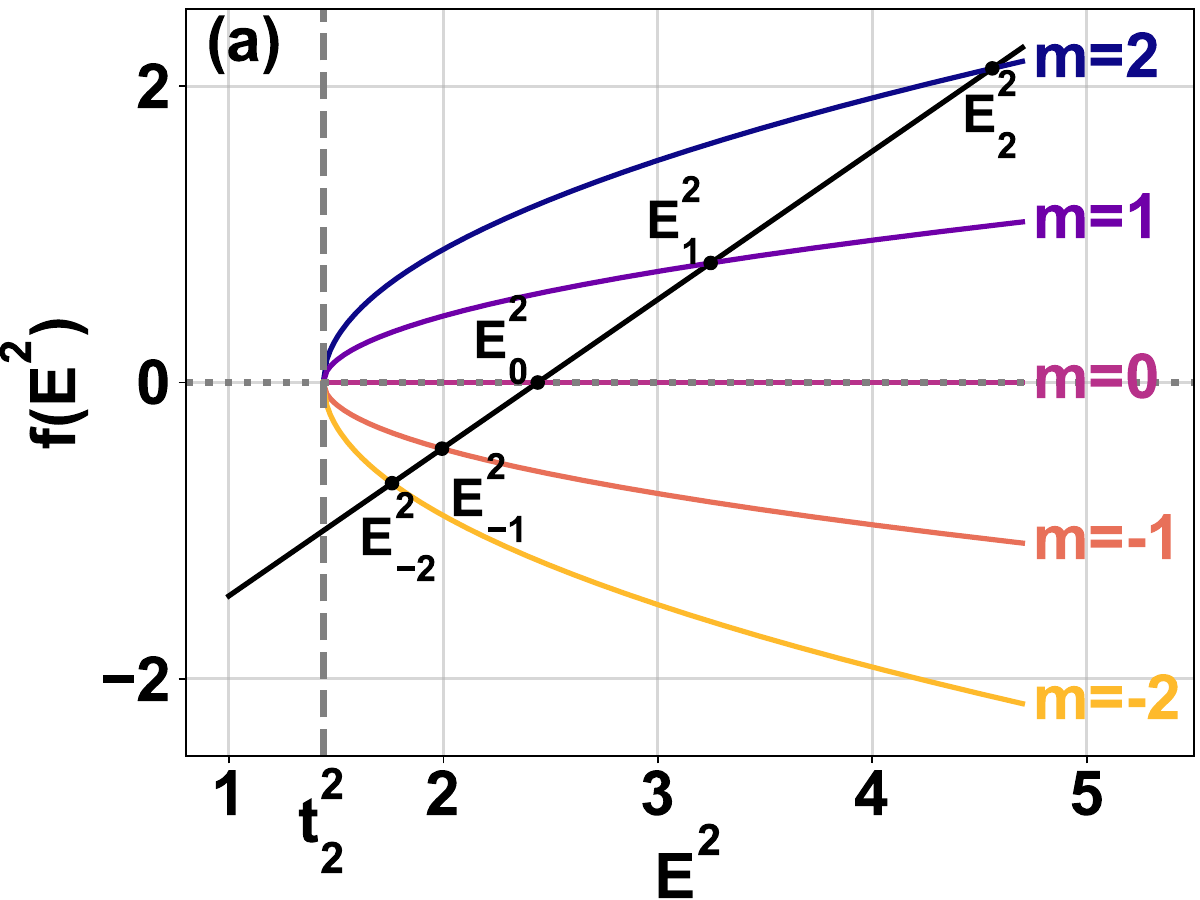}
    \end{minipage}
    \hfill
    \begin{minipage}[b]{0.49\linewidth}
        \centering
        \includegraphics[width=\linewidth]{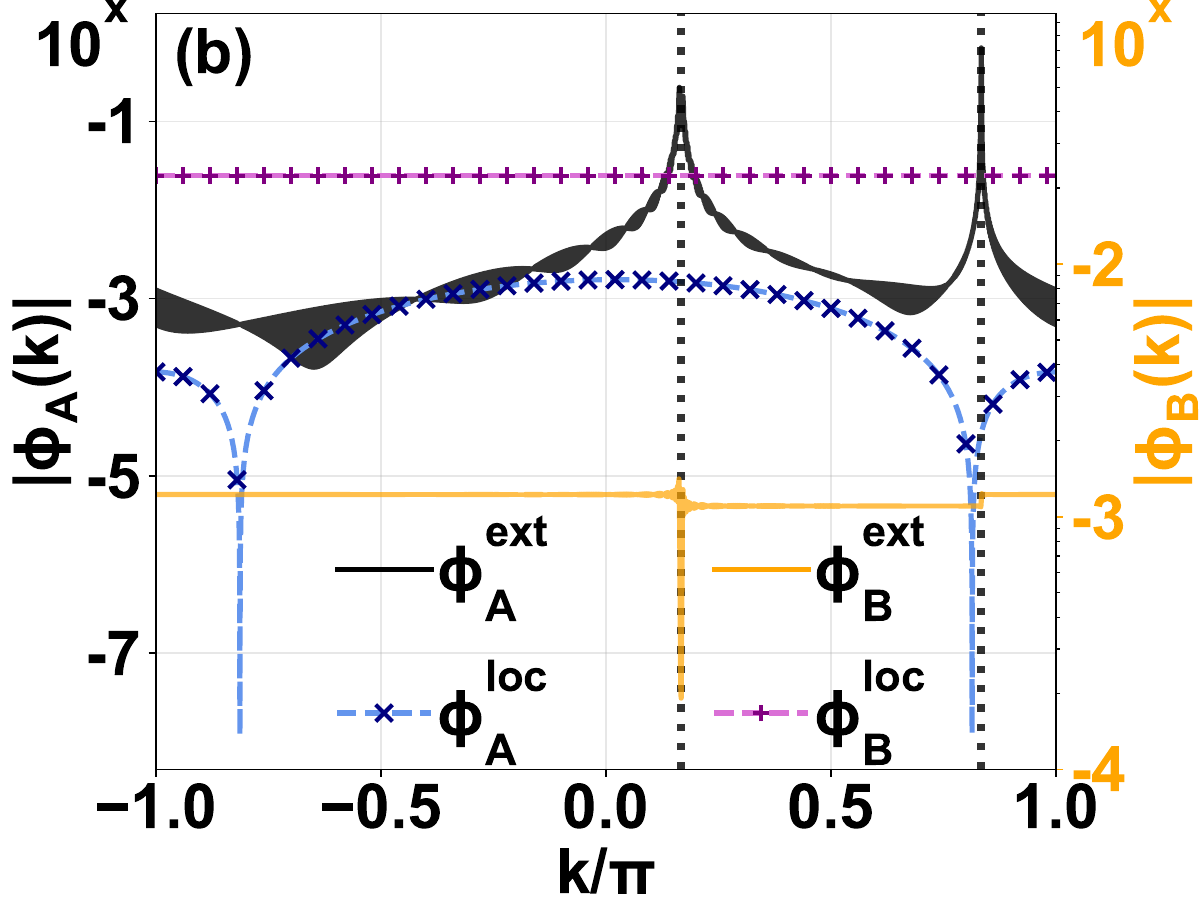}
    \end{minipage}

    \caption{
    Theoretical mechanism. (a) Graphical solution of the quantization condition. Black dots mark eigenenergies $E_m^2$ at the intersections of the constraint $f(E^2)=E^2-t_1^2-t_2^2$ (black line) and mode curves $y_m(E^2)$ (colored lines). Vertical dashed line indicates the threshold $t_2^2$. (b) Reciprocal space wavefunction amplitudes (logarithmic scale). Solid lines show the extended state ($E \approx 0.5 t_2$) for $A$ (black, left axis) and $B$ (orange, right axis) sublattices. Dashed lines (theoretical) and markers (numerical) plot the localized state ($m=50$). Vertical dotted lines mark the resonant momenta $k_c=\pm \arcsin (E/t_2)$. Other parameters match those in Fig.~1 [$L=2001$ ($N=1000$), $t_1=1$, $t_2=1.2$, and $F=0.6$].
    }
    \label{fig:theory_mechanism}
\end{figure}

\subsection{Localized Stark ladder ($|E| > t_2$)}
In the high-energy regime where $|E| > t_2$, the denominator $D(k) = E - t_2 \sin k$ is strictly nonzero for all $k \in [-\pi, \pi]$. The formal solution to Eq.~(\ref{eq:ODE}) is
\begin{equation}
\phi_B(k) = \mathcal{C} \exp\left[ -\frac{i}{F} \int^k \mathcal{V}(q, E) dq \right].
\end{equation}
Since $\mathcal{V}(q, E)$ is real-valued, the modulus $|\phi_B|$ is constant. The phase, however, accumulates non-trivially. Explicit integration yields the closed-form phase factor
\begin{equation}\label{eq:analytic}
\begin{aligned}
\arg [\phi_B(k)] &= \frac{E^2-t_1^2-t_2^2}{F t_2 \sqrt{1 - \alpha^2}} \ln\left( \frac{\alpha \tan (k/2) - 1 - \sqrt{1 - \alpha^2}}{\alpha \tan (k/2) - 1 + \sqrt{1 - \alpha^2}} \right) \\
&+ \frac{2t_1}{F} \ln(\alpha - \sin k),
\end{aligned}
\end{equation}
where $\alpha = E / t_2$. To satisfy the single-valuedness condition $\phi_B(k+2\pi) = \phi_B(k)$, the total phase accumulated across the Brillouin Zone (BZ) must be an integer multiple of $2\pi$. This leads to the quantization condition
\begin{equation}\label{eq:quantization}
\frac{E^2 - t_1^2 - t_2^2}{F \sqrt{E^2 - t_2^2}} = m, \quad m \in \mathbb{Z}.
\end{equation}
The quantization condition follows from the $2\pi$ phase winding of $\phi_B$ across the BZ, evaluated via the residue theorem (see Appendix~\ref{app:derivations}). This quantization condition is graphically solved in Fig.~\ref{fig:theory_mechanism}(a), where the intersections of the linear function $f(E^2)=E^2-t_1^2-t_2^2$ and the mode curves $y_m(E^2) = m F \sqrt{E^2-t_2^2}$ determine the discrete energy levels. Inverting Eq.~(\ref{eq:quantization}) yields the explicit energy spectrum
\begin{equation}\label{eq:E}
E_m^2 = t_1^2 + t_2^2 + \frac{mF}{2}\left( m F \pm \sqrt{m^2 F^2 + 4t_1^2} \right).
\end{equation}
This solution contains two distinct spectral branches, visualized in Fig.~\ref{fig:theory_mechanism}(a). These branches correspond to the blue (unbounded) and green (bounded) regions of the spectrum shown in Fig.~\ref{fig:main_results}(c). For large positive indices ($m \gg 1$), the energies scale as $E_m^\pm \approx \pm mF$, recovering the standard unbounded WS ladder (blue branch). Conversely, for large negative indices ($m \ll -1$), the energies are bounded and asymptotically approach the ME from above
\begin{equation}\label{eq:low_asymptotic}
E_m^2 \simeq t_2^2 + \frac{t_1^4}{m^2F^2} \quad \implies \quad |E_m| \to t_2^+.
\end{equation}
Thus, the localized spectrum consists of an infinite ladder extending to high energies and a dense accumulation of states (green branch) approaching $|E|=t_2$. The validity of our analytical derivation is further corroborated by the wavefunctions in reciprocal space; as shown in Fig.~\ref{fig:theory_mechanism}(b), the localized states (blue dashed lines and markers) remain smooth and bounded, in agreement with the theoretical prediction derived from Eq.~(\ref{eq:analytic}).

\subsection{\label{subsec:continuum} Extended continuum ($|E| < t_2$)}
Before analyzing the general solution in this regime, we consider a special case where the effect of the Stark potential is completely nullified~\cite{IGC_2024}: there exist specific momenta $k_0$ satisfying $t_1 + t_2 \cos k_0 = 0$. If the energy is tuned to $E^{\pm}(k_0) = t_2 \sin k_0 = \pm \sqrt{t_2^2 - t_1^2}$, the coupled equations admit a solution with a plane wave on the $A$ sublattice and no distribution on the $B$ sublattice. These two specific eigenstates are fully extended, surviving in the thermodynamic limit regardless of the potential strength $F$.

For general energies satisfying $|E| < t_2$ (but $E \neq E_0$), the solution is more intricate. The denominator $(E - t_2 \sin k)$ vanishes at two resonant momenta $k_c$ in the BZ, or equivalently, the two poles reside on the unit circle defined by $z=e^{ik}$. These poles partition the BZ into disjoint open intervals. Within each interval, the differential equation Eq.~(\ref{eq:ODE}) remains well defined, admitting local solutions of the form
\begin{equation}
\phi_B(k) = C_j \exp\left[ -\frac{i}{F} \int_{k_0}^k \mathcal{V}(q, E) dq \right],
\end{equation}
where $C_j$ is a constant specific to the $j$th interval. Expanding the denominator near a resonance, $\mathcal{V}(k) \approx \text{Res}(k_c)/(k-k_c)$, the phase integral develops a logarithmic divergence
\begin{equation}\label{eq:phi_b_phase}
\Theta(k) \sim -\frac{\text{Res}(k_c)}{F} \ln|k - k_c|,\quad \text{Res}(k_c) = -\frac{(t_1 + t_2 \cos k_c)^2}{t_2\cos k_c}.
\end{equation}
This implies that the wavefunction $\phi_B(k)$ possesses an essential phase singularity, oscillating infinitely rapidly as $k \to k_c$. Since the modulus $|\phi_B(k)|$ is determined by the imaginary part of $\mathcal{V}(k)$, which vanishes for real $E$, the amplitude of $\phi_B$ remains bounded almost everywhere. This piecewise constant behavior of $|\phi_B(k)|$ with abrupt jumps at the poles is consistent with the numerical results in Fig.~\ref{fig:theory_mechanism}(b) (green line).

The presence of these singularities fundamentally alters the quantization condition. In the high-energy regime, the periodicity of the BZ imposes a global phase closure constraint $\phi_B(\pi) = \phi_B(-\pi)$, which discretizes the energy. However, for $|E| < t_2$, the infinite phase accumulation at the poles effectively decouples the boundary conditions of the separate intervals. As a result, valid bounded solutions can be constructed for \textit{any} real energy $E$ in this range, giving rise to a continuous spectrum.

The nature of the eigenstates in this continuum is characterized by the $A$-sublattice component
\begin{equation}\label{eq:AB_relation}
\phi_A(k) = \frac{t_1 + t_2 \cos k}{E - t_2 \sin k} \phi_B(k).
\end{equation}
As $k \to k_c$, $\phi_A(k)$ inherits the phase singularity of $\phi_B(k)$ but is further modulated by the diverging denominator $(k-k_c)^{-1}$. Substituting this into Eq.~(\ref{eq:phi_b_phase}), we obtain the singularity structure of $\phi_A$ near a pole
\begin{equation}
    \phi_A(k) \sim  -|k-k_c|^{-i\xi - 1}\operatorname{sgn}(k-k_c), \quad \xi=\frac{\text{Res}(k_c)}{F}. 
\end{equation}
We evaluate the asymptotic behavior of the real-space wavefunction $\psi_{n,A}$ via the inverse Fourier transform. According to the asymptotic theory of Fourier transforms~\cite{Lighthill_1958}, the asymptotic behavior of the wavefunction is governed by the Fourier transform of these singularities
\begin{equation}
    \psi_{n,A} \propto \int dk \, \phi_A(k) e^{ikn} \sim \sum_{j=1}^2 A_j e^{ik_{c_j} n}|n|^{i\xi_j}\operatorname{sgn}(n),
\end{equation}
where $A_j$ are $n$-independent constants associated with the two poles. Since $\left| \lvert n \rvert^{i\xi_j}\right| = 1$, the amplitude $|\psi_{n,A}|$ remains non-decaying as $n\to\infty$, indicating an extended state. This behavior characterizes an AC spectrum, distinguishing our system from critical quasiperiodic models where states often exhibit the multifractal or self-similar structure of a singular continuous spectrum. This classification is supported by our numerical scaling analysis [Fig.~\ref{fig:localization_proof}(b)], which yields a fractal dimension $D_2 \approx 1.00$ over 6 orders of magnitude with sizes up to $L \sim 10^9$, suggesting the absence of the fractional scaling ($0 < D_2 < 1$) associated with critical states.

\section{Numerical Results}

We corroborate our theoretical findings with numerical diagonalization of the Hamiltonian under open boundary conditions (OBCs), examining the convergence to the thermodynamic limit. OBCs are chosen to avoid the large potential discontinuity at the boundaries inherent to Stark systems under periodic boundary conditions. While OBCs discretize the spectrum and introduce localized edge modes, the bulk density of states and the energy position of the ME are thermodynamic properties insensitive to boundary conditions~\cite{Pastur_1992}. Unless otherwise stated, we use the set of parameters $t_1=1$, $t_2=1.2$, $F=0.6$, $N=1000$. To quantify the localization properties, we utilize the IPR and the associated fractal dimension $D_2$~\cite{Thouless_1974, Evers_2008}, defined as
\begin{equation}
\text{IPR} = \sum_{n, \sigma} |\psi_{n,\sigma}|^4, \quad D_2 = -\frac{\ln(\text{IPR})}{\ln L},
\end{equation}
where $\sigma \in \{A, B\}$ denotes the sublattice index. For an extended state, $\text{IPR} \sim L^{-1}$ and $D_2 \to 1$, whereas for a localized state, $\text{IPR} \sim \text{const}$ and $D_2 \to 0$. We rely on these scaling indices rather than the Lyapunov exponent, as the latter is ill-defined for systems exhibiting hyperexponential WS localization~\cite{Longhi_2023}.

\begin{figure}[htbp]
    \centering
    \includegraphics[width=0.95\linewidth]{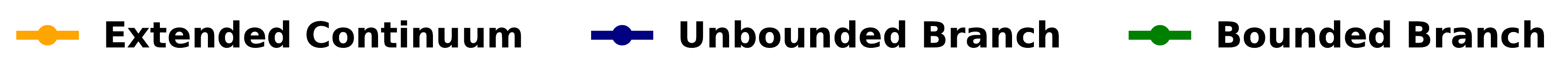}

    \begin{minipage}[b]{0.49\linewidth}
        \centering
        \includegraphics[width=\linewidth]{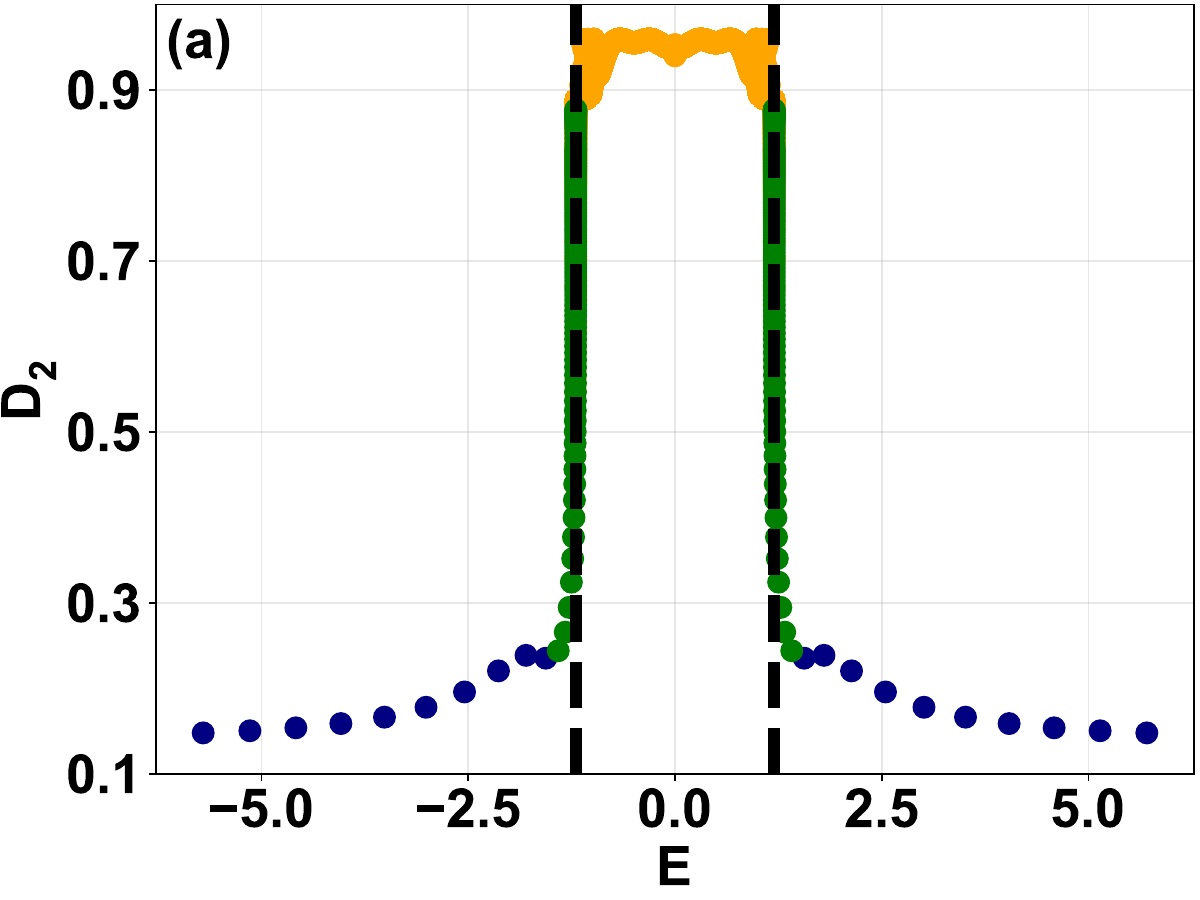}
    \end{minipage}
    \hfill 
    \begin{minipage}[b]{0.49\linewidth}
        \centering
        \includegraphics[width=\linewidth]{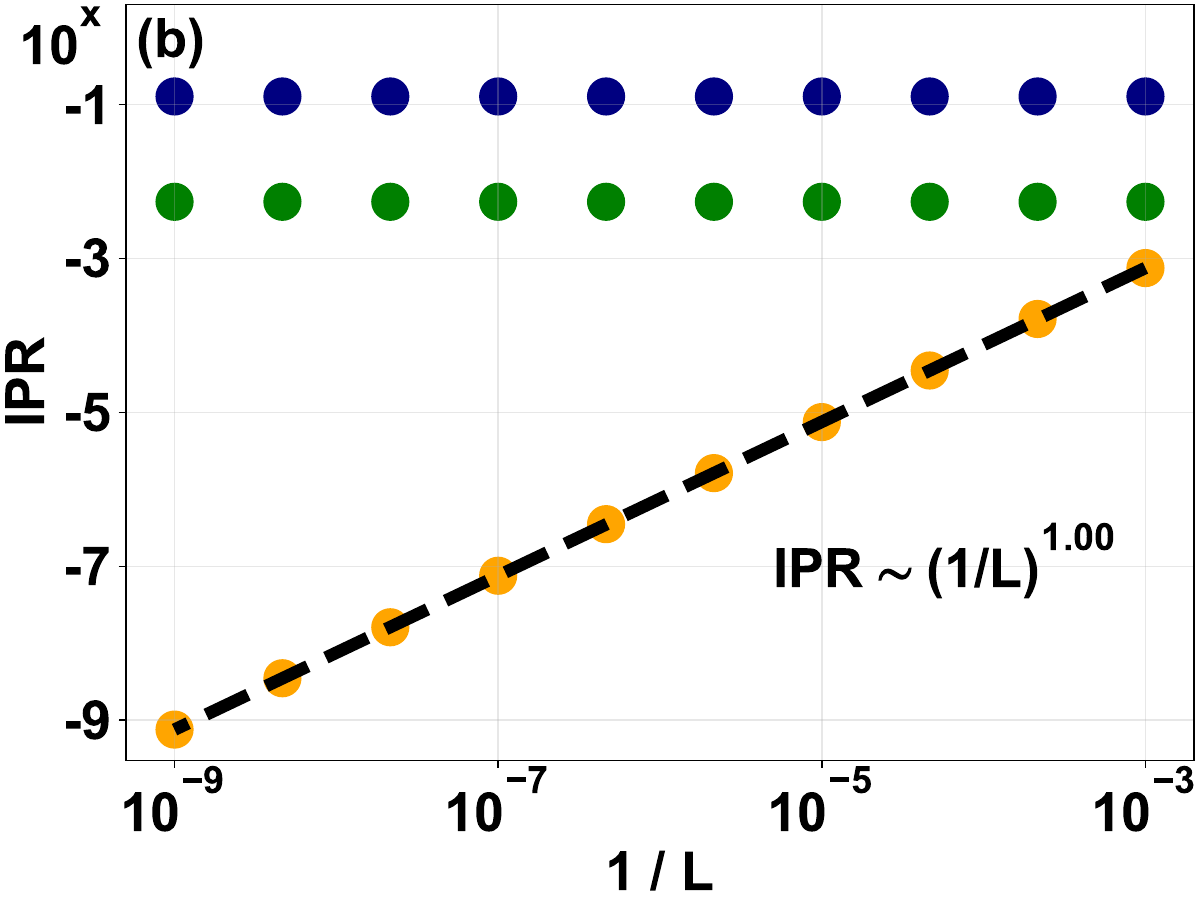}
    \end{minipage}
    
    \vspace{0.15cm} 
    
    \begin{minipage}[b]{0.49\linewidth}
        \centering
        \includegraphics[width=\linewidth]{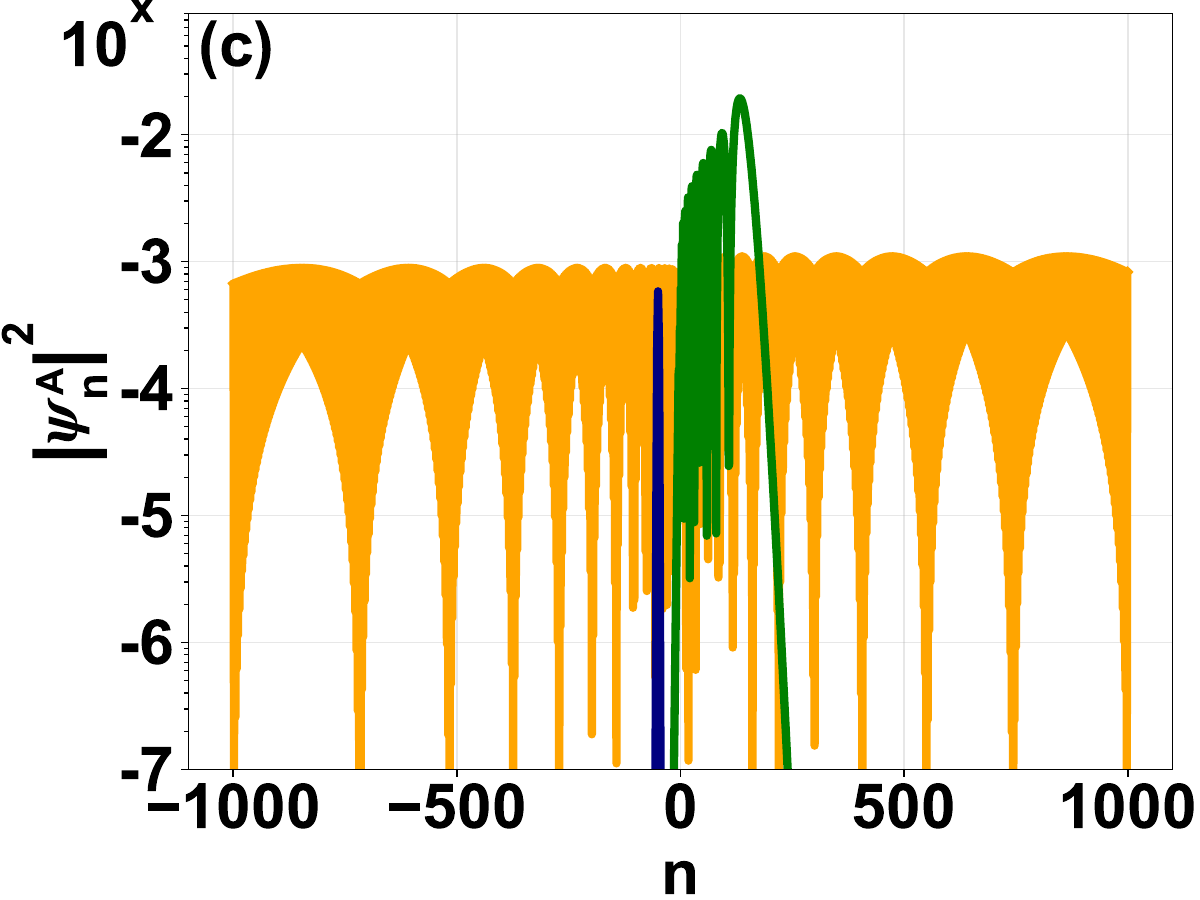}
    \end{minipage}
    \hfill
    \begin{minipage}[b]{0.49\linewidth}
        \centering
        \includegraphics[width=\linewidth]{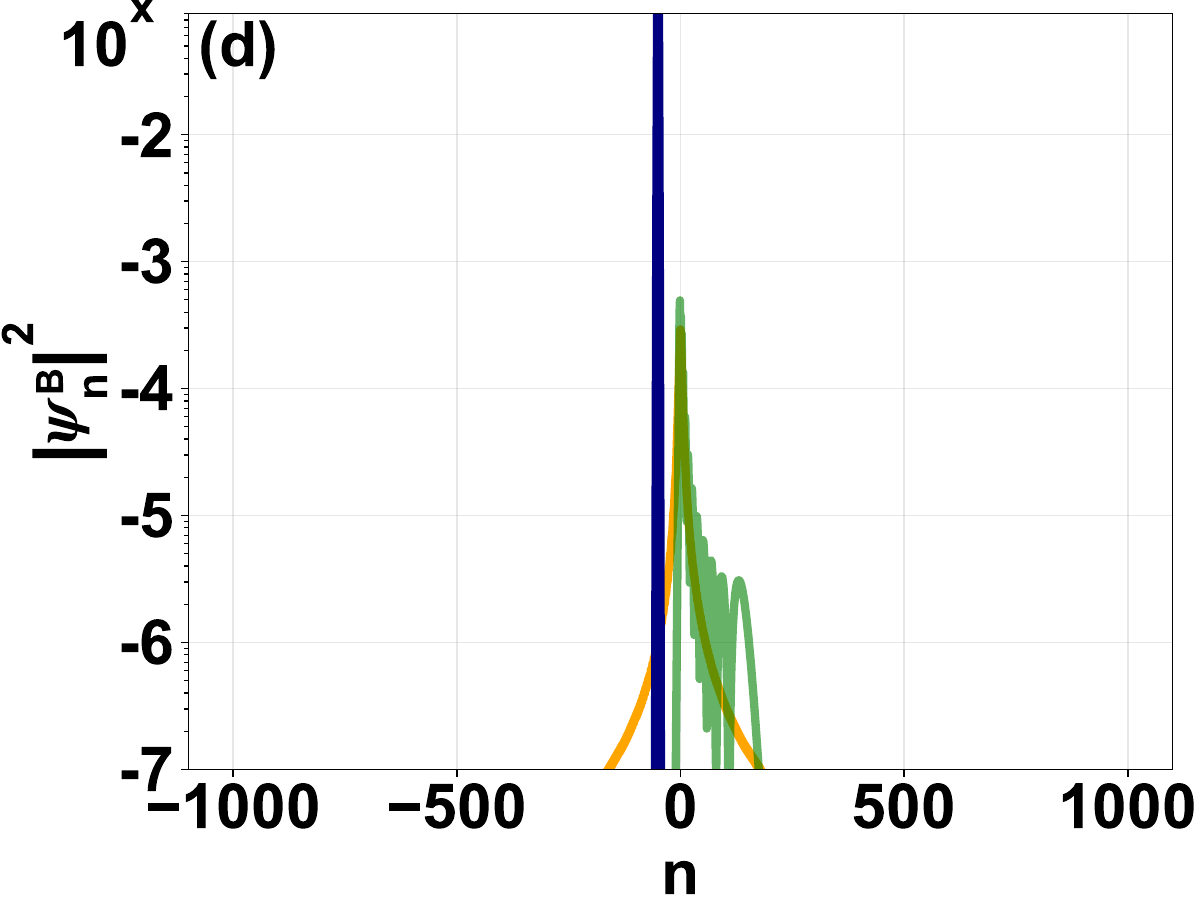}
    \end{minipage}

    \caption{
    Characterization of localization. (a) Fractal dimension ($D_2$) vs energy ($L=20001$, $N=10\ 000$, truncated at $|E| < 6t_1$). Black dashed lines mark the analytical mobility edges at $|E|=t_2$. (b) Finite-size scaling of inverse participation ratio (IPR) vs system size $L$ for representative states: extended continuum ($E \approx 0.5t_2$), unbounded branch ($m=50$), and bounded branch ($m=-10$). Dashed line indicates the scaling $\text{IPR} \sim L^{-1.00}$. (c),(d) Real-space probability density $|\psi_n|^2$ on $A$ and $B$ sublattices (logarithmic scale, truncated at $10^{-7}$) shows the spatial distribution for the representative states ($L=2001, N=1000$). Other parameters match those in Fig.~1 ($t_1=1$, $t_2=1.2$, and $F=0.6$).
}
    \label{fig:localization_proof}
\end{figure}

The global phase diagram is presented in Fig.~\ref{fig:main_results}(d). To examine the sharpness of the transition, we plot $D_2$ as a function of energy for a large system size ($N=10\ 000$) for the whole spectrum in Fig.~\ref{fig:localization_proof}(a). A distinct steplike transition is observed at the analytical ME $|E|=t_2$. States within the continuum ($|E| < t_2$, red) exhibit $D_2 \approx 1$, indicating delocalization. In the localized regime ($|E| > t_2$), two distinct behaviors appear: the unbounded branch ($m>0$, blue) shows $D_2 \approx 0$ as expected, while the bounded branch ($m<0$, green) exhibits intermediate fractal dimensions ($0.2 \lesssim D_2 \lesssim 0.8$). This elevated $D_2$ arises from the branch's accumulation near the critical point, making it difficult to distinguish from critical or extended states using a fixed system size.

To resolve this ambiguity and address the broader challenge of distinguishing true extended states from finite-size artifacts (``pseudo-MEs''), we rely on finite-size scaling. As demonstrated in the context of mosaic Stark lattices~\cite{Longhi_2023}, states that appear extended at moderate sizes ($L \sim 10^4$) may eventually localize at extremely large scales ($L > 10^7$). To determine the behavior in the thermodynamic limit of both the continuum and the anomalous bounded branch, we perform a finite-size scaling analysis of the IPR up to $N = 5 \times 10^8$ (total unit cells $L = 2N + 1 \approx  10^9$, total sites $2L \approx 2 \times 10^9$). We provide more details about the large-scale calculations in Appendix~\ref{app:calculation} and the robustness check in Appendix~\ref{app:robustness}.

Figure~\ref{fig:localization_proof}(b) displays the IPR scaling for representative states from each branch. The extended continuum state (orange) follows a power law $\text{IPR} \propto L^{-1.00}$ across six orders of magnitude without deviation, providing evidence that the delocalization is not a transient effect but a persistent feature of the thermodynamic limit. In contrast, the bounded branch (green), despite residing close to the ME, exhibits a saturation of IPR ($\text{slope} \approx 0$), consistent with its localized nature. The lower saturation value of the IPR for the bounded branch compared to the unbounded branch reflects a larger localization length, which is consistent with its proximity to the delocalization transition.

Finally, we visualize the spatial structure of these states in Figs.~\ref{fig:localization_proof}(c) and~\ref{fig:localization_proof}(d). The wavefunction distributions reveal the distinct mechanisms governing each branch. The unbounded branch (blue) is highly concentrated on the $B$ sublattice, directly pinned by the Stark potential. The bounded branch (green) resides primarily on the $A$ sublattice, allowing it to minimize the potential energy penalty and spread over a larger (but finite) range. The extended state (orange) is dominated by the $A$ sublattice, where it forms a nondecaying standing wave, while its weight on the $B$ sublattice is significantly suppressed, consistent with the asymptotic behavior predicted by our analytical derivation.

\section{Experimental Proposal and Stability}
\label{sec:experiment}

We propose an experimental realization of our model using photonic frequency synthetic dimensions on a TFLN platform. This approach leverages the high tunability of electro-optic modulation to engineer complex hopping phases and site-dependent potentials, as recently demonstrated for the Creutz ladder topology~\cite{PhotonicCreutz_2025} and Bloch oscillations~\cite{BlochOscillation_2016}.

\subsection{Setup}
\begin{figure}[htbp]
    \centering
    \begin{minipage}[b]{0.49\linewidth}
        \centering
        \includegraphics[width=\linewidth]{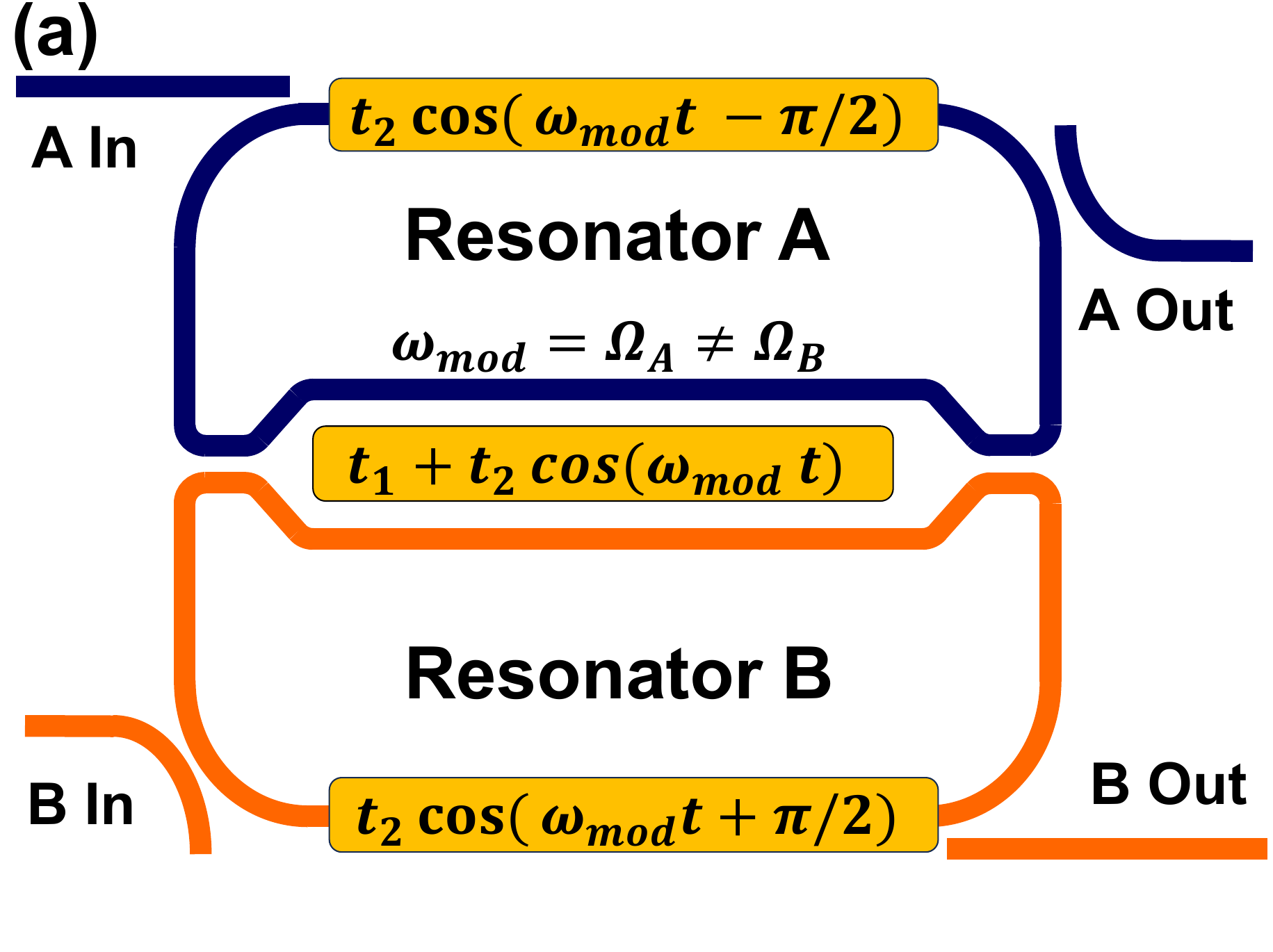}
    \end{minipage}
    \hfill 
    \begin{minipage}[b]{0.49\linewidth}
        \centering
        \includegraphics[width=\linewidth]{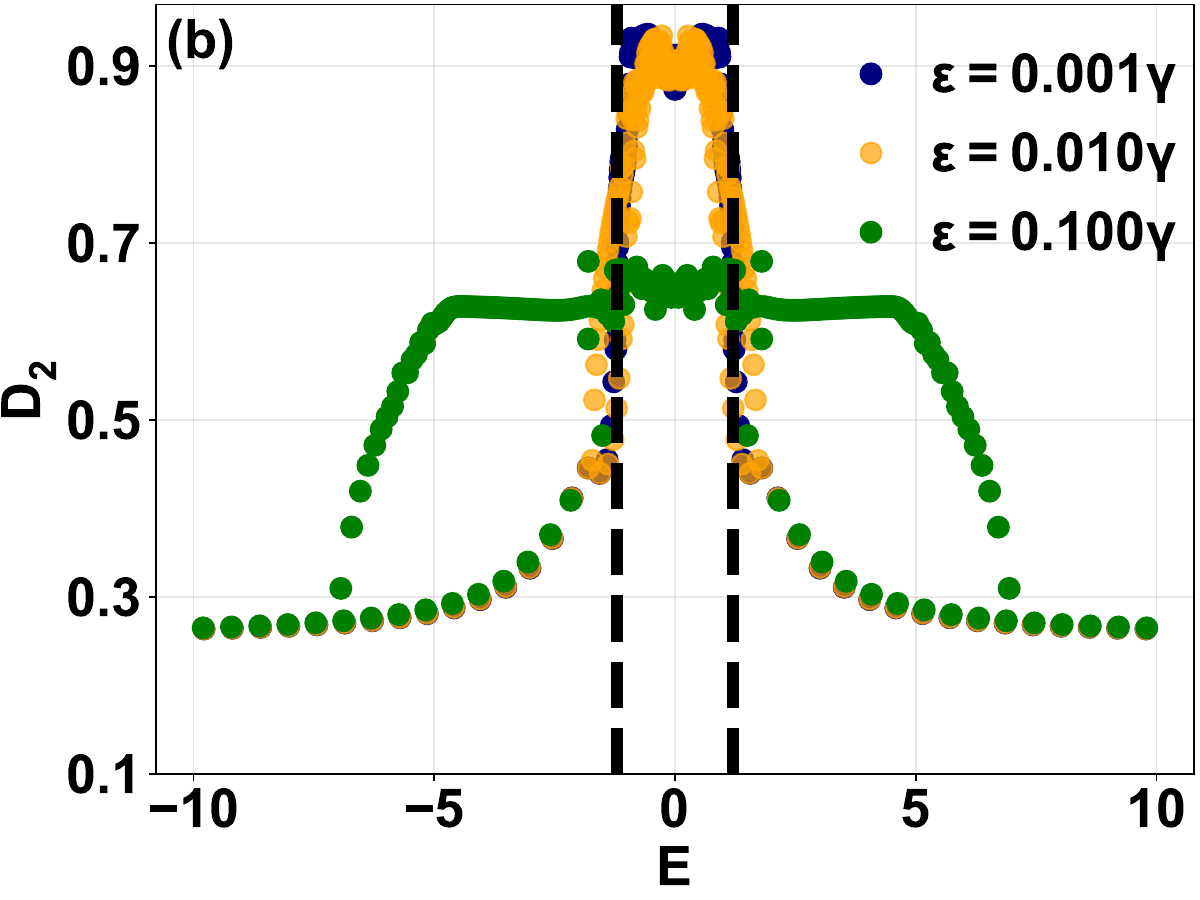}
    \end{minipage}
    
    \vspace{0.15cm} 
    
    \begin{minipage}[b]{0.49\linewidth}
        \centering
        \includegraphics[width=\linewidth]{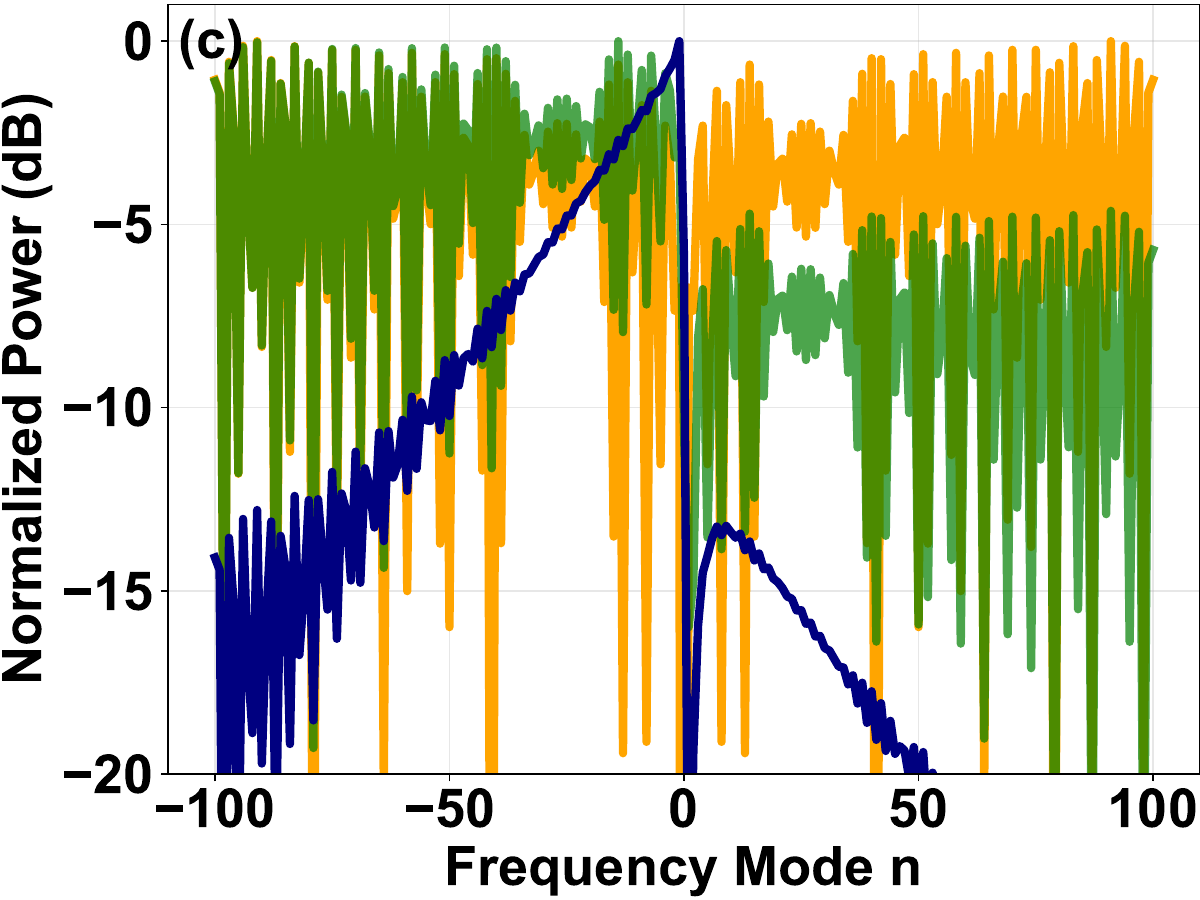}
    \end{minipage}
    \hfill
    \begin{minipage}[b]{0.49\linewidth}
        \centering
        \includegraphics[width=\linewidth]{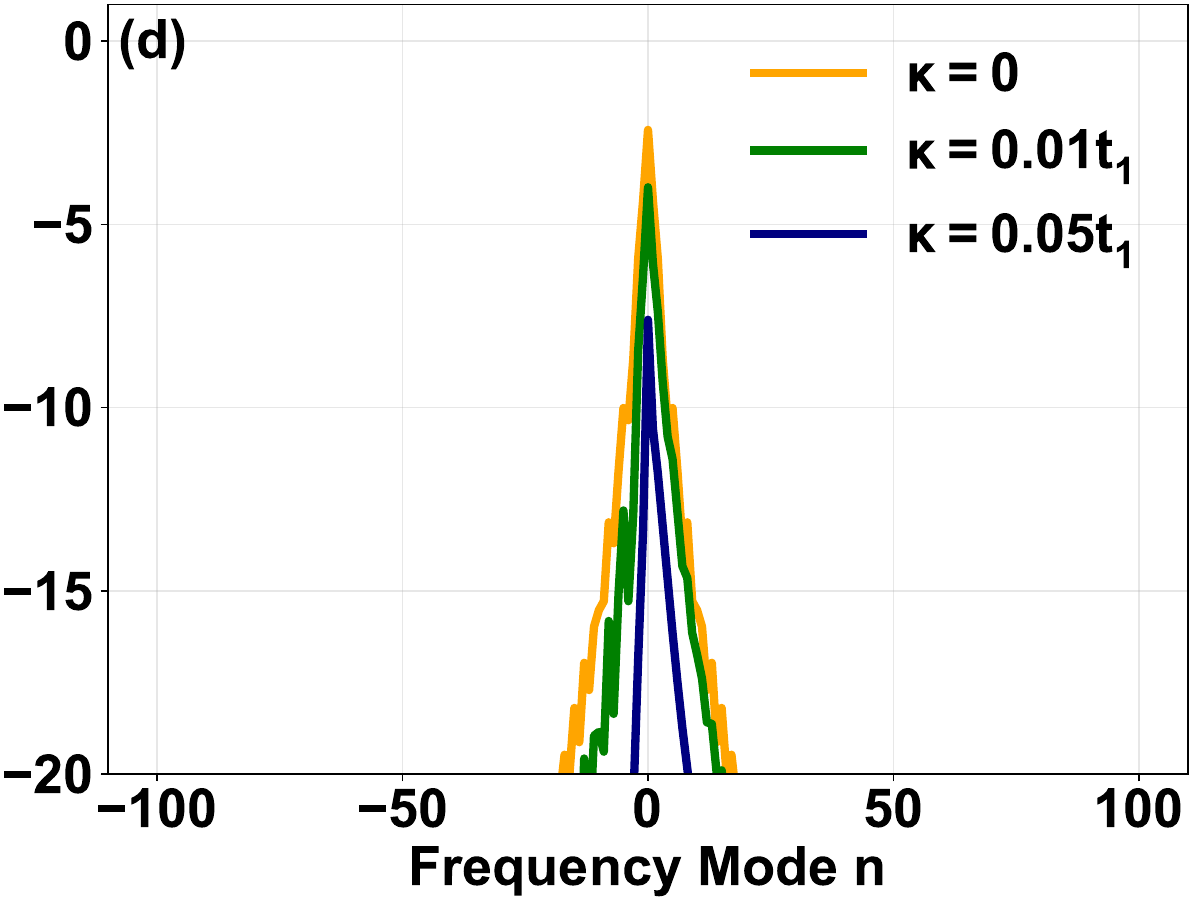}
    \end{minipage}
    \caption{
    Experimental proposal and robustness analysis.
    (a) Schematic of the proposed photonic platform. Two ring resonators (A and B) are coupled via a tunable Mach-Zehnder Interferometer (MZI). RF electrodes on the rings generate intraresonator hopping ($\pm it_2/2$), while the MZI generates interresonator couplings ($t_1, t_2/2$). A DC bias on resonator B detunes its FSR to tune the Stark potential gradient $F$.
    (b) Robustness of the fractal dimension ($D_2$) against leakage potential strength $\epsilon$ on the $A$ sublattice with $\epsilon/F \in\{0.001, 0.01, 0.1\}$ (system size $L=201, N=100$). The energy is truncated by $|E| < 10$.
    (c, d) Simulated steady-state output spectra with leakage potential $\epsilon=0.01F$ and different loss rates $\kappa/t_1\in\{0, 0.01, 0.05\}$ from resonator A (c) and resonator B (d) under injection at the center mode of resonator A ($E=0$). The spectra are normalized to the maximum intensity at port A. Other parameters match those in Fig.~1 ($t_1=1$, $t_2=1.2$, and $F=0.6$).
    }
    \label{fig:exp}
\end{figure}

The experimental setup, illustrated in Fig.~\ref{fig:exp}(a), consists of two coupled ring resonators, A and B, which represent the two legs of the ladder. The lattice sites correspond to the frequency modes of the resonators, spaced by the FSR. As proposed in Ref.~\cite{PhotonicCreutz_2025}, the complex hoppings required for the Creutz ladder topology can be realized through a combination of intraresonator and interresonator modulations. Specifically, electro-optic modulators (EOMs) on each resonator driven at the FSR frequency generate the horizontal hopping terms ($A_n \leftrightarrow A_{n+1}, B_n \leftrightarrow B_{n+1}$) with phases $\pm i t_2/2$ controlled by the radio frequency (rf) drive phase. The vertical hopping ($t_1$) and cross hopping ($t_2/2$) are realized by connecting the resonators via a tunable Mach-Zehnder Interferometer (MZI) modulated by both direct current (DC) and RF signals~\cite{PhotonicCreutz_2025}.

The crucial ingredient of our model—a linear Stark potential acting exclusively on the $B$ sublattice—is implemented by detuning the FSR of resonator B relative to the modulation frequency. In the frame rotating at the modulation frequency $\omega_{\text{mod}}$, the effective on-site potential for mode $n$ is given by $V_n = n (\Omega_{\text{FSR}} - \omega_{\text{mod}})$~\cite{BlochOscillation_2016}. By lithographically designing resonator B to have a slightly different FSR from resonator A ($\Omega_\text{B} \neq \Omega_\text{A}$) and locking the modulation frequency to resonator A ($\omega_{\text{mod}} = \Omega_\text{A}$), we achieve a flat potential on the $A$ sublattice ($V_n^\text{A} \approx 0$) and a linear gradient on the $B$ sublattice ($V_n^\text{B} = F n$), where $F = \Omega_\text{B} - \Omega_\text{A}$. The FSR can be fine-tuned via a DC bias voltage on the EOM, allowing precise control of the Stark potential parameter. We provide more details on the derivation of effective Hamiltonian in photonic systems in Appendix~\ref{app:rotation_frame}. 

\subsection{Robustness analysis}
Realizing this proposal requires addressing three key experimental imperfections: carrier frequency mismatch, potential leakage, and photon loss.

First, the unavoidable mismatch in the central resonant frequencies of the two resonators introduces a constant energy offset $\delta_{\text{BA}}$ between the $A$ and $B$ sublattices, described by $H_{\text{offset}} = \sum_n \delta_{\text{BA}} b_n^\dagger b_n$. This term shifts the zero-energy reference of the $B$ leg [i.e. $\mathcal{V}(k,E)\to\mathcal{V}(k,E)+\delta_{\text{BA}}$ in Eq.~(\ref{eq:ODE})], adding a phase factor $\exp(-i\delta k/F)$ to the $B$ sublattice wavefunction in momentum space. Crucially, this constant shift does not alter the localization properties or the existence of the ME, as the quantization condition for discrete energies and the no-pole condition remains invariant.

Second, imperfect locking of the modulation frequency to the FSR of resonator A introduces a residual ``leakage'' potential $V_n^\text{A} = \epsilon n$. While any nonzero $\epsilon$ theoretically localizes the extended states in the thermodynamic limit, we must assess their stability within the finite bandwidth of realistic photonic lattices (several hundreds of modes~\cite{FrequencyComb_2019}). We simulate the fractal dimension ($D_2$) of the eigenstates for a system size $L=201$ ($N=100$) under a leakage potential $H_{\text{leak}} = \sum_n \epsilon n a^\dagger_n a_n$. As shown in Fig.~\ref{fig:exp}(b), the ME remains distinguishable and the continuum states retain their extended character ($D_2 \approx 1$) for leakage ratios $\epsilon/F \lesssim 10^{-2}$. This indicates that frequency locking stability on the order of 1\% of the Stark gradient is sufficient for observation.

Third, the finite photon lifetime $\kappa^{-1}$ competes with the hopping timescales, damping the extended states before they can propagate across the frequency lattice. The experimental observable is the steady-state transmission spectrum, which corresponds to the Green's function of the system. For a laser injection at frequency $\Omega_0$ into the central mode of resonator A, the output field vector $\mathbf{s}_{\text{out}}$ is given by (see Appendix~\ref{app:input_output})
\begin{equation}
    \mathbf{s}_{\text{out}} \propto i(H - \Omega_0 - i\kappa/2)^{-1} \mathbf{s}_{\text{in}},
\end{equation}
where $\mathbf{s}_{\text{in}}$ is localized at the injection site. We simulate the output spectrum for an injection energy $\Omega_0=0$ (center of the extended band) in Figs.~\ref{fig:exp}(c) and~\ref{fig:exp}(d). The results show that for low loss ($\kappa \le 0.01 t_1$), the spectrum exhibits a broad, flat distribution on the $A$ sublattice (theoretically the $A$-sublattice distribution dominates), characteristic of an extended state. However, as the loss increases to $\kappa = 0.05 t_1$, the spectrum becomes asymmetric and decays away from the injection site, limiting the effective propagation range. This highlights the necessity of the strong coupling regime ($\kappa \ll t_1$).

\subsection{Feasibility estimation}
We adopt realistic parameters reported in state-of-the-art TFLN literature to perform a numerical feasibility study. High-$Q$ resonators ($Q \sim 1.5 \times 10^6$) with strong electro-optic coupling have been demonstrated~\cite{FrequencyComb_2019}, achieving coupling strengths comparable to the FSR ($t_1 \sim 0.6 \times \text{FSR}$). For a typical FSR of $2\pi \times 10$ GHz, this yields a hopping strength $t_1 \approx 2\pi \times 5$ GHz. Our model parameter $F = 0.6 t_1$ corresponds to a detuning of $\sim 2\pi \times 3$ GHz (30\% of FSR). To avoid weakening the coupling efficiency due to large detuning~\cite{BlochOscillation_2016}, we suggest operating with a smaller gradient (e.g., $F \sim 2\pi \times 1$ GHz, 10\% of FSR), which remains well within the tunable range. Under this detuning, the $1\%$ potential leakage threshold requires a frequency locking accuracy of $2\pi \times 10$ MHz.

The primary challenge is the loss rate. An intrinsic $Q$-factor of $1.5 \times 10^6$ at 1550 nm corresponds to an intrinsic decay rate $\kappa_0 \approx 2\pi \times 130$ MHz. By operating in the undercoupled regime to minimize broadening from external coupling, the total linewidth $\kappa$ can be maintained close to this intrinsic limit. With $t_1 \approx 2\pi \times 5$ GHz, the ratio $\kappa/t_1 \approx 2.6\%$, which is slightly above our conservative threshold of 1\%. Some potential approaches to further decrease this ratio include utilizing ultralow loss TFLN fabrication techniques ($Q\sim 10^7$~\cite{HighQ_2017, HighQ_2024}) or by increasing the FSR and RF drive efficiency~\cite{LowV_2019, EnhanceCoupling_2025} to boost the effective hopping strength. Thus, the observation of ME signatures is feasible with current or near-future TFLN technology.

\section{Conclusion}
In summary, we have proposed and solved a disorder-free dimerized Stark lattice hosting an exact ME. By mapping the system to an unbounded Jacobi matrix, we demonstrate that the model evades the restrictions of the Simon-Spencer theorem, enabling the existence of true extended states in the presence of an unbounded potential. Analytically, we identify the transition at $|E|=t_2$, separating a continuum of extended states from two distinct localized branches: a standard Wannier--Stark ladder and an anomalous bounded branch accumulating at the ME. These predictions are corroborated by large-scale finite-size scaling ($L \sim 10^9$), which supports the persistence of the extended phase in the thermodynamic limit. 

Furthermore, we propose an experimental realization using photonic frequency synthetic dimensions on TFLN resonators. We perform a stability analysis simulating experimental imperfections based on realistic parameters. Our results suggest that the ME is robust against FSR mismatch and photon loss, provided the system operates in the strong coupling regime. Our work not only resolves the theoretical ambiguity surrounding MEs in Stark systems but also establishes a general mechanism for realizing and observing delocalization in the presence of unbounded potentials using modern photonic platforms.

\begin{acknowledgments}
The authors thank Peijie Chang and Dan Long for their helpful discussions. This work is supported by the National Natural Science Foundation of China under Grant No.~62131002.
\end{acknowledgments}

\appendix
\section{Gauge Transformation to Real Symmetric Form}
\label{app:gauge_transformation}
In this appendix, we detail the unitary transformation that maps the effective Hamiltonian to a real symmetric tridiagonal form (a Jacobi matrix).

Recall that the effective model derived in the main text possesses a real intercell hopping $J_{\text{inter}} = t_2$ and a complex intracell hopping
\begin{equation}
J_{\text{intra}}^{\pm}(n)= t_1\pm iF n/2 = \rho_n e^{\pm i\theta_n},
\end{equation}
where the magnitude is $\rho_n = \sqrt{t_1^2+F^2 n^2/4}$ and the phase is $\theta_n = \arctan(\frac{F n}{2t_1})$.

We define a site-dependent gauge transformation via the unitary operator $S$, which acts on the basis states as $|A_n\rangle \to e^{i\Phi_{n,A}}|A_n\rangle$ and $|B_n\rangle \to e^{i\Phi_{n,B}}|B_n\rangle$. To render the intracell hopping real, the phase difference within a cell must satisfy $\Phi_{n,B} - \Phi_{n,A} = \theta_n$. Simultaneously, to keep the intercell hopping $t_2$ (connecting $B_n$ and $A_{n+1}$) real, we require $\Phi_{n+1,A} = \Phi_{n,B}$.

These recursion relations are satisfied by the cumulative sums
\begin{equation}
    \Phi_{n,A} = \sum_{k=-N}^{n-1} \theta_k, \quad \Phi_{n,B} = \sum_{k=-N}^{n} \theta_k.
\end{equation}
The corresponding transformation matrix takes the form
\begin{equation}
    S = \text{diag}(e^{-i\Phi_{-N,A}}, e^{-i\Phi_{-N,B}}, \cdots, e^{-i\Phi_{N,A}}, e^{-i\Phi_{N,B}}).
\end{equation}
Applying the transformation $H'' = S^{-1} H' S$, the complex phases are eliminated. The resulting Hamiltonian $H''$ is a real symmetric tridiagonal matrix with off-diagonal entries
\begin{equation}
    J^{\text{real}}_{\text{intra}}(n) = \sqrt{t_1^2+\frac{F^2 n^2}{4}}, \quad J^{\text{real}}_{\text{inter}} = t_2,
\end{equation}
while the on-site potential $V_n = -F n/2$ remains invariant. This demonstrates that the system maps to a Jacobi matrix where a subsequence of the off-diagonal terms grows unbounded, matching the growth of the diagonal potential.

\section{Restrictions on the Spectrum of Jacobi Matrices}
\label{app:Simon_Spencer}

We first cite Theorem 2.1 in the original work of Simon and Spencer~\cite{SimonSpencer_1989}, which forbids the AC spectrum with unbounded potentials. In Theorem 2.1, they assume the operator $h$ to be
\begin{equation}\label{eq:SS_assumption}
    (hu)(n) = u(n+1)+u(n-1)+v(n)u(n)
\end{equation}
on $l^2 (\mathbb{Z})$. Then the unbounded potential
\begin{equation}
    \limsup_{n\to\infty} |v(n)| = \limsup_{n\to -\infty} |v(n)| = \infty
\end{equation}
leads to an empty AC spectrum. From Eq.~(\ref{eq:SS_assumption}) we clearly see the hopping term is assumed to be NN and uniform [$u(n\pm 1)$ has coefficient 1]. Our effective model violates the uniformity condition ($a_n \sim n$). Furthermore, our original model Eq.~(\ref{eq:model}) violates the NN condition directly due to the sublattice spinor structure. Thus, the $n$-dependent hopping helps evade the restriction from the Simon-Spencer theorem.

Next, we examine broader conditions for the existence or absence of spectral components in unbounded Jacobi matrices in mathematical physics. We will demonstrate that our effective Hamiltonian $H''$ does not satisfy the sufficient conditions for an empty AC spectrum  or empty pure point (PP) spectrum found in the relevant literature~\cite{Cojuhari_2008, Swiderski_2016}.

We adopt the standard notation for Jacobi matrices, where $\{a_n\}_{n=0,1,\cdots}$ denotes the off-diagonal weights (hopping) and $\{b_n\}_{n=1,2,\cdots}$ denotes the diagonal terms (potential). The standard mathematical formulation assumes a semi-infinite lattice with boundary condition $a_0=0$. This half-line setup is equivalent to considering the $n>0$ and $n<0$ halves of our system separately; it does not affect the existence of the AC spectrum in the bulk. Mapping our model to the index $j \in \mathbb{N}$ (where $2n+1 \to A_n$ and $2n+2\to B_n$), the entries are
\begin{equation}\label{eq:entries}
\begin{aligned}
    a_{2n} &= t_2, \quad &a_{2n+1} &= \sqrt{t_1^2+F^2 n^2/4}, \\
    b_{2n} &= -\frac{F n}{2}, \quad &b_{2n+1} &= -\frac{F n}{2}.
\end{aligned}
\end{equation}

Theorem 2.2(c) of Ref.~\cite{Cojuhari_2008} provides a sufficient condition for the \textit{absence} of an AC spectrum. The spectrum is empty if
\begin{equation}
\begin{aligned}
    \liminf_{n\to\infty} a_n > 0, &\quad \sup_n \frac{b_n^2+b_{n+1}^2}{a_n^2} < \infty, \\
    \liminf_{n\to \infty}&\left(1-\frac{b_n b_{n+1}}{a_n^2}\right) > 0.
\end{aligned}
\end{equation}
Applying Eq.~(\ref{eq:entries}) to the subsequence where $a_k = t_2$ (the even terms $a_{2n}$), we find
\begin{equation}
    \frac{b_{2n}^2+b_{2n+1}^2}{a_{2n}^2} = \frac{(F n/2)^2 + (F n/2)^2}{t_2^2} = \frac{F^2 n^2}{2t_2^2} \to \infty.
\end{equation}
Since the supremum is infinite, the condition is not satisfied. Thus, this theorem does not prohibit the AC spectrum in our model.

Regarding the PP spectrum, Ref.~[\citealp{Swiderski_2016}, Theorem 1.1] provides conditions forbidding the PP spectrum of a Jacobi matrix. A necessary condition for this theorem to hold is
\begin{equation}
    \limsup_{n\to\infty} \frac{|b_n|}{a_n} < 2.
\end{equation}
For our model, taking the subsequence $a_{2n}=t_2$, we have $|b_{2n}|/a_{2n} \propto n/t_2 \to \infty$. Thus, the theorem does not apply, and the existence of a PP spectrum (our localized Stark ladder) is allowed.

In summary, the staggered nature of the weights $\{a_n\}$—alternating between bounded ($t_2$) and unbounded ($J_{\text{intra}} \sim n$)—is the critical feature. It places the model in a ``gap'' between theorems that require uniform or bounded hopping (Simon-Spencer) and theorems designed for uniformly growing matrices. We direct readers interested in the mathematical construction of other operators supporting MEs to Refs.~[\citealp{Janas_2003},~\citealp{Janas_2009}].

\section{Derivation of the Spectrum}
\label{app:derivations}

Here we explicitly calculate the phase integral of the potential $\mathcal{V}(k, E)$ to obtain the analytic expression for the wavefunction phase. We decompose the integral into two parts
\begin{equation}\label{eq:integral_decompose}
\begin{aligned}
I(k) &= \int^k \frac{t_1^2 + t_2^2 + 2t_1t_2 \cos q - E^2}{E - t_2 \sin q} dq \\
&= (t_1^2 + t_2^2 - E^2)\int^k \frac{dq}{E - t_2 \sin q} + \int^k \frac{2t_1t_2 \cos q}{E - t_2 \sin q} dq.
\end{aligned}
\end{equation}
The second term is the integral of an exact differential, yielding $-2t_1 \ln (\alpha - \sin k)$ with $\alpha = E / t_2$. The first term is solved using the standard Weierstrass substitution $x = \tan (q /2)$. The integral transforms as
\begin{equation}
\begin{aligned}
\int \frac{dq}{\alpha - \sin q} &= \int \frac{2 dx}{\alpha(1+x^2) - 2x} \\ 
&=\frac{1}{\sqrt{1 - \alpha^2}} \ln \left( \frac{\alpha \tan (k/2) - 1 - \sqrt{1 - \alpha^2}}{\alpha \tan (k/2) - 1 + \sqrt{1 - \alpha^2}} \right).
\end{aligned}
\end{equation}
Combining these results and exponentiating gives the closed-form wavefunction
\begin{widetext}
\begin{equation}\label{eq:phi_b}
    \phi_B(k) = \left(\frac{\alpha \tan \frac{k}{2} - 1 - \sqrt{1 - \alpha^2}}{\alpha \tan \frac{k}{2} - 1 + \sqrt{1 - \alpha^2}}\right)^{\frac{i(E^2-t_1^2-t_2^2)}{F t_2 \sqrt{1 - \alpha^2}}} \cdot (\alpha - \sin k)^{2it_1 / F}.
\end{equation}
\end{widetext}
This corresponds to Eq.~(\ref{eq:analytic}) in the main text. 
To ensure the wavefunction is single valued in the BZ, we impose the condition $\phi_B(\pi) = \phi_B(-\pi)$. While the function is periodic, the logarithmic term in Eq.~(\ref{eq:phi_b}) accumulates a phase of $2\pi$ when traversing the BZ. Specifically, the argument of the logarithm has two branches; the branch with the positive imaginary part spans the range $-\pi$ to $\pi$, while the negative branch spans the range $2\pi$ to $\pi$. This $2\pi$ phase winding enforces the exponent to be an integer. Note that for localized states $|E|>t_2$, the term $\sqrt{1-\alpha^2}$ becomes purely imaginary, rendering the prefactor of the logarithm real
\begin{equation}\label{eq:app_quantization}
\frac{E^2 - t_1^2 - t_2^2}{F t_2 \sqrt{\alpha^2-1}} = m, \quad m \in \mathbb{Z}.
\end{equation}
Substituting $\alpha = E/t_2$ recovers the condition provided in Eq.~(\ref{eq:quantization}). 
Here we introduce another approach to achieve this result via the residue theorem. We calculate the quantization condition directly on the unit circle $z = e^{ik}$. The phase integral is
\begin{equation}
I = \oint_{|z|=1} \frac{t_1^2 + t_2^2 - E^2}{E - \frac{t_2}{2i}(z - z^{-1})} \frac{dz}{iz}.
\end{equation}
Note that the integral of the cosine term [from Eq.~(\ref{eq:integral_decompose})] vanishes because it is the integral of an exact differential $\frac{d}{dk}\ln(E-t_2\sin k)$ over a closed period. The poles of the remaining integrand are roots of $t_2z^2 - 2iEz - t_2 = 0$
\begin{equation}
z_{\pm} = \frac{i(E \pm \sqrt{E^2 - t_2^2})}{t_2}.
\end{equation}
Assuming $|E| > t_2$, only one root (denoted $z_1$) lies inside the unit circle, as $z_+ z_- = -1$. Applying the residue theorem gives $I = 2\pi ( E^2 - t_1^2 - t_2^2)/\sqrt{E^2 - t_2^2}$. The single-valuedness condition requires $\exp(-iI/F) = 1$ leading to
\begin{equation}
\frac{E^2 - t_1^2 - t_2^2}{F \sqrt{E^2 - t_2^2}} = m, \quad m \in \mathbb{Z},
\end{equation}
which is the same as Eq.~(\ref{eq:app_quantization}). 

Solving for $E$ yields the explicit energy levels
\begin{equation}
E_m^2 = t_1^2 + t_2^2 + \frac{mF}{2}\left( m F + \sqrt{m^2 F^2 + 4t_1^2} \right).
\label{eq:E_herm_app}
\end{equation}
Note that we have chosen the positive sign for the square root; taking $m$ to cover all integers $\mathbb{Z}$ generates both spectral branches. 

We verify the consistency of this solution with the ``no-pole'' assumption ($|E| > t_2$). For positive $m$, $E_m^2$ is monotonically increasing and unbounded, clearly satisfying the condition. For negative $m$, we analyze the monotonicity by taking the derivative with respect to $m$
\begin{equation}
\frac{\partial E_m^2}{\partial m} = mF^2 \left(1 - \frac{m^2F^2 / 2 + t_1^2}{\sqrt{m^4 F^4 / 4 + m^2t_1^2F^2}}\right).
\end{equation}
Since $(m^2F^2 / 2 + t_1^2)^2 = m^4 F^4 / 4 + m^2t_1^2F^2 + t_1^4$, the term in parentheses is negative. Since $m < 0$, the prefactor is negative, making the total derivative positive. Thus, $E_m^2$ increases with $m$ (decreases as $m \to -\infty$). The asymptotic limit is
\begin{equation}
E_m^2 \xrightarrow{m \to -\infty} t_2^2 + \frac{t_1^4}{m^2F^2}.
\end{equation}
Since the asymptotic value approaches $t_2^2$ from above, all solutions satisfy $|E| > t_2$, demonstrating their validity.

\section{Large-Scale Numerical Methods}
\label{app:calculation}

To distinguish the true MEs from finite-size effects, we perform numerical diagonalization on system sizes up to $L \approx 10^9$. Standard sparse diagonalization techniques (e.g., ARPACK) typically encounter memory bottlenecks or 32-bit integer indexing limits at $L \sim 10^8$. To overcome these limitations, we implement a custom shift-invert Krylov subspace algorithm optimized for the specific structure of our Hamiltonian. Since the rotation and transformation are local and the gauge transformation does not change the localization property, we can use the tridiagonal real symmetric matrix $H^{\prime\prime}$ derived in Appendix~\ref{app:gauge_transformation} to do the large-scale calculation, which can reduce the memory requirement by restricting to real numbers and utilizing the tridiagonal form.

The method relies on three key optimizations:
\begin{enumerate}
    \item \textbf{Shift-invert spectral transformation:} To target interior eigenvalues (e.g., the continuum states near $E=0.5t_2$), we solve the transformed eigenvalue problem $(H - \sigma I)^{-1} \psi = \nu \psi$, where $\sigma$ is the target energy shift~\cite{Saad_2011}. The state closest to $\sigma$ corresponds to the eigenvalue $\nu$ with the largest magnitude, which converges rapidly in the Krylov iteration. To track the evolution of specific states, we set $\sigma = E_m^{\text{theo}}$ for discrete localized levels (fixed index $m$) and $\sigma = 0.5t_2$ for the extended continuum, ensuring we probe bulk properties away from edge effects.
    
    \item \textbf{Banded linear solver:} Under OBCs, the effective Hamiltonian $H^{\prime\prime}$ (derived via the gauge transformation) is tridiagonal. Instead of general sparse formats, we store $H$ in a packed banded format. The linear inversion step $(H - \sigma I)x = b$ is performed using the Thomas algorithm (via LAPACK's \texttt{gtsv} routine), which requires only $O(L)$ time and memory, avoiding the fill-in associated with general sparse $LU$ decomposition.
    
    \item \textbf{Disk-backed Krylov subspace:} For $L \sim 10^9$, storing the Krylov basis vectors exceeds the RAM capacity of standard computing nodes. We utilize memory mapping (\texttt{np.memmap}) to offload the basis vectors to high-speed NVMe storage. This allows the simulation to scale linearly with disk space rather than RAM.
\end{enumerate}

We employ the Lanczos algorithm using double precision ($\texttt{np.float64}$), which requires storing only two vectors in memory for orthogonalization, allowing us to resolve the fine spectral structure of the system up to $L \sim 10^9$.

\section{Algorithm Validation and Stability}
\label{app:robustness}
To ensure the reliability of our custom implementation, we perform benchmarks against standard libraries and statistical verification.

\subsection{Accuracy comparison} 
We validate our disk-backed solver against the standard \texttt{scipy.sparse.linalg.eigsh} implementation for system sizes where the latter fits in memory ($L\approx10^3$ to $10^6$). Figure~\ref{fig:benchmark}(a) shows the relative error of energy $|\Delta E|/|E|$ for the three representative states used in the main text. We repeat the calculation 10 times and report the maximum relative error in this figure. For the discrete levels, the error is close to machine precision for both branches while it is more significant (with maximum value $\sim 10^{-13}$) for the continuum band. The larger relative error observed in the continuum branch is a signature of the extremely dense spectrum, where the eigenvalue conditioning deteriorates rapidly with size and results in numerical instability of the solution. Note that an offset of $10^{-20}$ was added to the plot to handle exact zero differences on the logarithmic scale.
\setcounter{figure}{0}
\renewcommand{\thefigure}{E\arabic{figure}}
\begin{figure}[htbp]
    \centering
    \begin{minipage}[b]{0.49\linewidth}
        \centering
        \includegraphics[width=\linewidth]{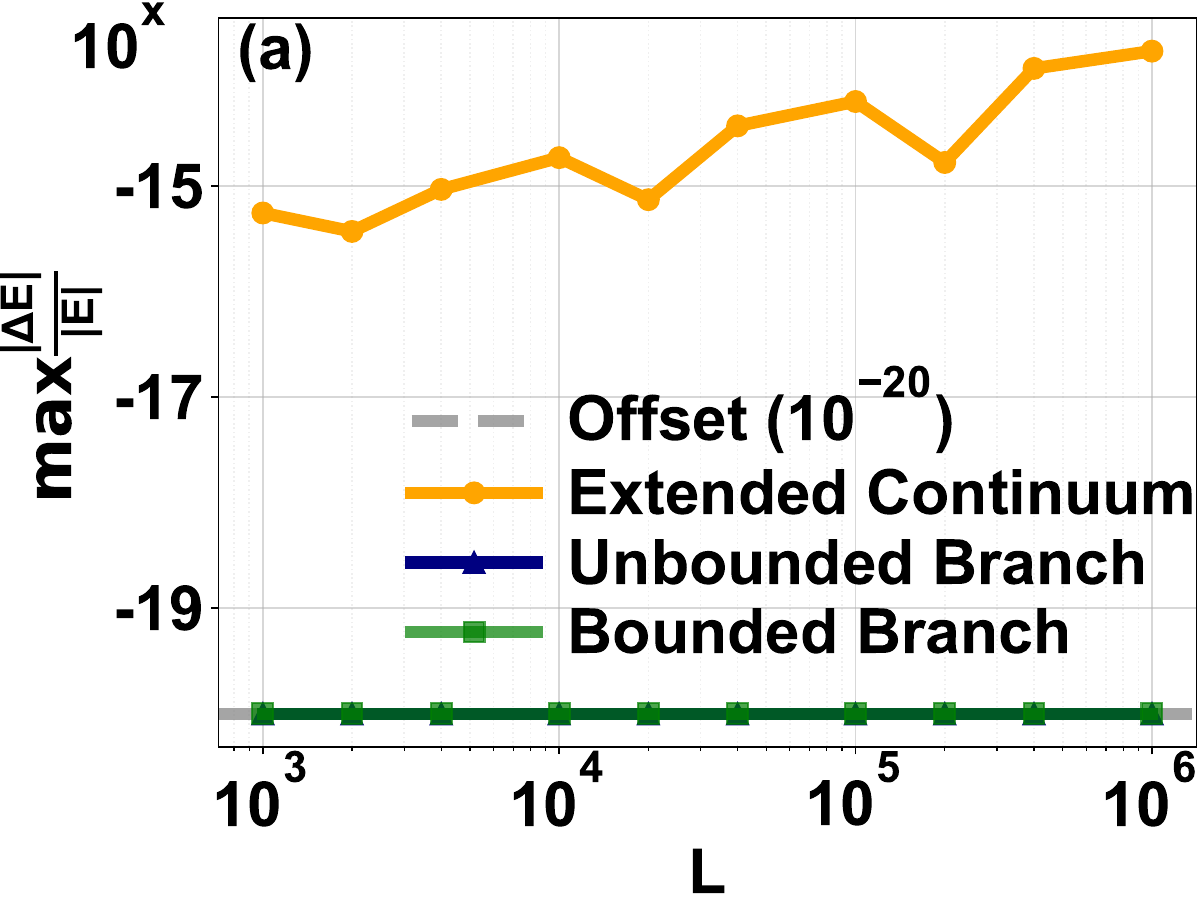}
    \end{minipage}
    \hfill
    \begin{minipage}[b]{0.49\linewidth}
        \centering
        \includegraphics[width=\linewidth]{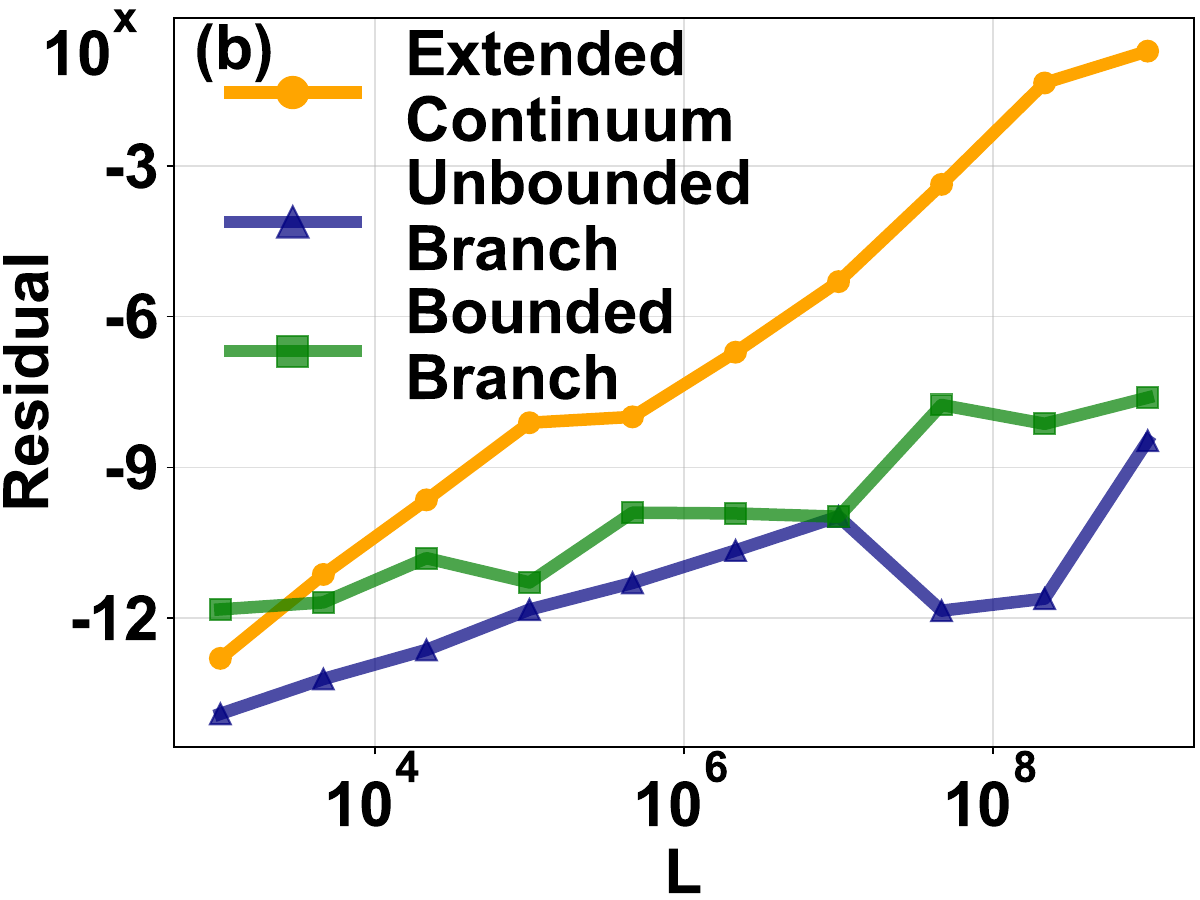}
    \end{minipage}
    
    \vspace{0.2cm}
    
    \caption{
        Maximum relative energy error compared to standard scipy solvers (calculated as the maximum over 10 independent runs per point) for $L\approx10^3$ to $10^6$. Plots include a $10^{-20}$ offset. (b) Residual $R = ||(H-E)\psi||/||\psi||$ vs system size $L$ for $L\approx 10^3$ to $10^9$.
    }
    \label{fig:benchmark}
\end{figure}

\subsection{Residuals} 
We further monitor the residual $R = ||(H - E)\psi|| / ||\psi||$ for the results in Fig.~\ref{fig:localization_proof}(c) (system sizes up to $L \sim 10^9$). As shown in Fig.~\ref{fig:benchmark}(b), the residual for discrete localized states grows slowly and remains relatively small ($< 10^{-7}$), demonstrating the accuracy of the eigenpairs. In contrast, the residual for the continuum branch grows with $L$ quickly. This growth is physically significant: it arises because the level spacing in the continuum vanishes as $L\to\infty$, causing the condition number of the shift-invert operator to diverge. Thus, the increasing residual serves as indirect support for the formation of a continuous spectrum in the thermodynamic limit.

\subsection{Statistical universality} 
\begin{figure}[htbp]
    \centering
        \includegraphics[width=\linewidth]{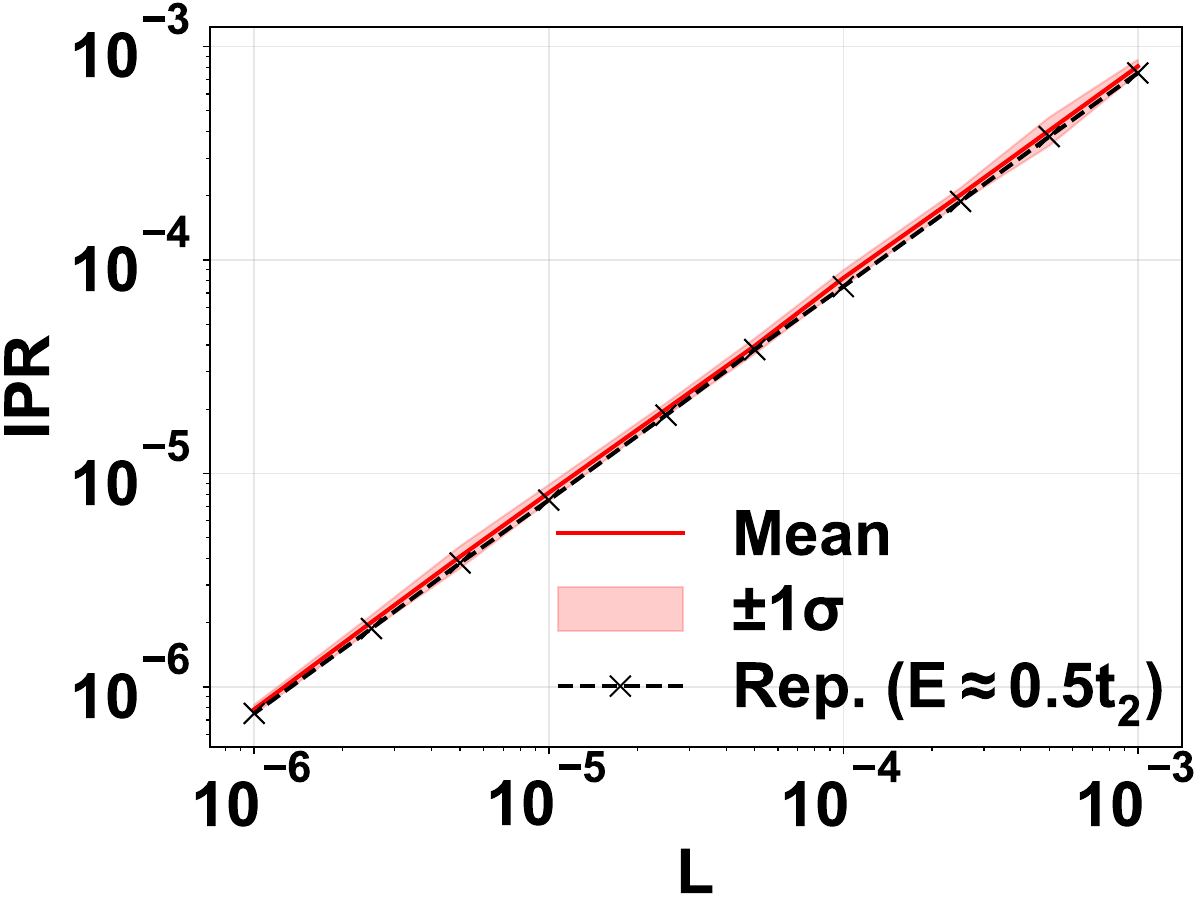}
    \caption{
        Statistical robustness of inverse participation ratio (IPR) scaling for continuum states. Solid lines show the mean IPR of 20 random states sampled from $0.2t_2\le|E| \le 0.8t_2$. Shaded regions indicate the standard deviation.
    }
    \label{fig:statistics}
\end{figure}

To verify that the single state tracked in the main text ($E \approx 0.5t_2$) is representative of the entire extended phase, we calculate the IPR for 20 randomly sampled states within the extended continuum for each system size. Figure~\ref{fig:statistics} displays the mean IPR (red line) and the standard deviation (shaded region). The variance is minimal, and the IPR of the representative state lies within the $\pm \sigma$ band. This suggests that the $L^{-1}$ scaling is a universal bulk property of the continuum, robust against microscopic energy variations.

\section{Derivation of the Effective Photonic Hamiltonian and Measurement Theory}
\label{app:exp}

In this appendix, we provide the detailed derivation of the effective static Hamiltonian from the time-dependent electro-optic modulation and establish the connection between the steady-state transmission spectrum and the Green's function calculated in the main text.

\subsection{Rotating frame transformation}\label{app:rotation_frame}
Consider a ring resonator supporting a set of frequency modes. In the absence of group velocity dispersion, the eigenfrequencies are equidistant: $\omega_n = \omega_0 + n\Omega_{\text{FSR}}$, where $n$ is the mode index relative to the carrier frequency $\omega_0$, and $\Omega_{\text{FSR}}$ is the FSR. The bare Hamiltonian in the laboratory frame is
\begin{equation}
    H_{0} = \sum_n (\omega_{\text{A}} + n\Omega_{\text{A}}) a_n^\dagger a_n + (\omega_{\text{B}} + n\Omega_{\text{B}}) b_n^\dagger b_n,
\end{equation}
where $\omega_{A(B)}$ and $\Omega_{A(B)}$ are the central frequency and FSR of resonator A(B), respectively. Electro-optic modulation at frequency $\omega_{\text{mod}}$ modulates the refractive index, introducing coupling between different frequency modes (horizontal coupling). The time-dependent modulation Hamiltonian reads
\begin{equation}
\begin{aligned}
    H_{h}(t) = 2t_{h} \cos(\omega_{\text{mod}}t - \pi/2) \sum_{n,m} (a_{n+m}^\dagger a_n + a_n^\dagger a_{n+m})\\
    + 2t_{h} \cos(\omega_{\text{mod}}t + \pi/2) \sum_{n,m} (b_{n+m}^\dagger b_n + b^\dagger_n b_{n+m}),
\end{aligned}
\end{equation}
where $t_{h}$ is the hopping strength proportional to the RF drive amplitude. We apply two coherent RF drives with the same hopping strength on the two resonators and a $\pi$ phase difference, yielding the $\pm\pi/2$ in the phases. Similarly, the DC bias and RF drive on the MZI contribute to the coupling between the two resonators. The DC bias couples modes with the same frequencies (vertical coupling)
\begin{equation}
    H_{v} = t_v\sum_n (a^\dagger_nb_n + b^\dagger_na_n),
\end{equation}
and the RF term couples modes with different frequencies (cross coupling)
\begin{equation}
    H_{c}(t) = 2t_{c}\cos(\omega_{\text{mod}}t) \sum_{n,m} (a_{n+m}^\dagger b_n + b^\dagger_n a_{n+m}).
\end{equation}
Note that the RF drive on the MZI originates from the same coherent source as the modulation on the individual resonators, ensuring the frequency is identical and the phase is locked.

To eliminate the time dependence, we transform to a frame rotating at the modulation frequency using the unitary operator
\begin{equation}
    U(t) = \exp\left[-i t \sum_n (\omega_{\text{A}} + n\omega_{\text{mod}})(a_n^\dagger a_n + b_n^\dagger b_n)\right].
\end{equation}
The effective Hamiltonian governing the dynamics in the rotating frame is given by $H_{\text{eff}} = U^\dagger (H_{0} + H_{h} + H_{v} + H_{c}) U + i \dot{U}^\dagger U$.

The time derivative term contributes a diagonal correction
\begin{equation}
    i \dot{U}^\dagger U = -\sum_n\left[ (\omega_{\text{A}} + n\omega_{\text{mod}}) (a_n^\dagger a_n +b^\dagger_nb_n)\right].
\end{equation}
Combining this with the bare Hamiltonian $U^\dagger H_{0} U = H_{0}$, the net diagonal term becomes
\begin{equation}
\begin{aligned}
    H_{\text{diag}} &= \sum_n n\left[ (\Omega_{\text{A}}-\omega_{\text{mod}}) a_n^\dagger a_n + (\Omega_{\text{B}}-\omega_{\text{mod}}) b_n^\dagger b_n \right] \\
    &+\sum_n (\omega_{\text{B}}-\omega_{\text{A}})b^\dagger_n b_n.
\end{aligned}
\end{equation}
This derivation explicitly shows that detuning the resonator's FSR from the modulation frequency generates a linear Stark potential $V_n = \Delta n$ while the difference in the central frequencies leads to a constant shift on the B-sublattice. We define the detuning parameters $\Delta_{\text{A\(B\)}} = \Omega_{\text{A\(B\)}} - \omega_{\text{mod}}$ and $\delta_{\text{BA}} = \omega_{\text{B}}-\omega_{\text{A}}$. By designing slightly different FSRs for the two resonators, we set $\Delta_\text{A} \approx 0$ (resonance) and $\Delta_\text{B} =F\neq 0$ (detuned) to realize the sublattice-selective potential.

For the off-diagonal terms, the operator transformation is 
\begin{equation}
\begin{aligned}
    U^\dagger a_{n+m}^\dagger a_n U &= U_{m+n}^\dagger a_{n+m}^\dagger U_{m+n} U_n^\dagger a_n U_n \\
    &=e^{im\omega_{\text{mod}}t} a_{n+m}^\dagger a_n
    , 
\end{aligned}
\end{equation}
where $U_n = \exp \left[-i t (\omega_{\text{A}} + n\omega_{\text{mod}})(a_n^\dagger a_n + b_n^\dagger b_n)\right]$. Similarly, we have 
\begin{equation}
\begin{aligned}
    U^\dagger b_{n+m}^\dagger a_n U =
    e^{im\omega_{\text{mod}}t} b_{n+m}^\dagger a_n, \\
    U^\dagger a_{n+m}^\dagger b_n U =
    e^{im\omega_{\text{mod}}t} a_{n+m}^\dagger b_n.
\end{aligned}
\end{equation}
Expanding the cosine modulation as $\left[e^{i(\omega_{\text{mod}}t + \phi)} + e^{-i(\omega_{\text{mod}}t + \phi)}\right]/2$, and applying the rotating wave approximation to discard the rapidly oscillating terms, we obtain the static coupling
\begin{equation}
\begin{aligned}
    U^\dagger H_hU &\approx it_h \sum_n (a_{n+1}^\dagger a_n + b_n^\dagger b_{n+1}- a_{n}^\dagger a_{n+1} - b_{n+1}^\dagger b_n), \\
    U^\dagger H_c U &\approx t_c \sum_n (a^\dagger_{n+1}b_n + b^\dagger_na_{n+1} + b_{n+1}^\dagger a_n + a^\dagger_n b_{n+1}) , \\
    U^\dagger H_vU &= H_v.
\end{aligned}
\end{equation}
Thus, the Hamiltonian under time-dependent modulation maps to a static lattice in the frequency dimension in the rotating frame, realizing the model described by Eq.~(\ref{eq:model}) with the addition of the potential shift $\sum_n \delta_{\text{BA}} b^\dagger_n b_n$ .

\subsection{Input-output formalism and transmission spectrum}
\label{app:input_output}
To connect the theoretical Hamiltonian to the experimental transmission signal, we employ the Heisenberg-Langevin formalism~\cite{Walls_2008}. The time evolution of the annihilation operators for the cavity modes, collected in the vector $\mathbf{\Psi} = (\dots, a_n, \dots, b_n, \dots)^T$, is given in the Heisenberg picture by
\begin{equation}\label{eq:langevin}
    \frac{d\mathbf{\Psi}}{dt} = -i [ \mathbf{\Psi}, H_{\text{lab}}(t) ] - \frac{\kappa}{2} \mathbf{\Psi} + \sqrt{\kappa_{\text{ex}}} \mathbf{S}_{\text{in}}(t),
\end{equation}
where $H_{\text{lab}} = H_0+H_v+H_c+H_h$ is the time-dependent laboratory frame Hamiltonian, $\kappa$ is the total loss rate, $\kappa_{\text{ex}}$ is the external coupling rate, and $\mathbf{S}_{\text{in}}(t)$ is the input noise vector. To remove the time dependence, we transform to the rotating frame defined by the operators
\begin{equation}
    \tilde{a}_n = a_n e^{i(\omega_{\text{A}}+n\omega_{\text{mod}})t}, \quad 
    \tilde{b}_n = b_n e^{i(\omega_{\text{A}}+n\omega_{\text{mod}})t}.
\end{equation}
Defining the rotating-frame vector $\mathbf{\Phi} = (\dots, \tilde{a}_n, \dots, \tilde{b}_n, \dots)^T$ and substituting it into Eq.~(\ref{eq:langevin}), the time derivatives of the exponential factors generate diagonal detuning terms. These terms combine with the modulation terms to yield a time-independent equation of motion
\begin{equation}
    \frac{d\mathbf{\Phi}}{dt} = -i H_{\text{eff}} \mathbf{\Phi} - \frac{\kappa}{2}\mathbf{\Phi} +\sqrt{\kappa_{\text{ex}}}\tilde{\mathbf{S}}_{\text{in}}(t),
\end{equation}
where $H_{\text{eff}}$ is the static matrix representation of the effective Hamiltonian derived in Appendix~\ref{app:rotation_frame}.

We consider a coherent continuous-wave laser input into the waveguide coupled to resonator A. In the laboratory frame, the input field at frequency $\omega_{\text{laser}}$ targeting the central mode ($n=0$) is $s_{\text{in}}^{\text{lab}}(t) = \mathcal{E} e^{-i\omega_{\text{laser}} t}$. The effective drive term in the rotating frame is
\begin{equation}
    s_{\text{in}}^{\text{rot}}(t) = s_{\text{in}}^{\text{lab}}(t) e^{i\omega_{\text{A}} t} = \mathcal{E} e^{-i(\omega_{\text{laser}} - \omega_{\text{A}})t}.
\end{equation}
By locking the laser frequency to the central resonance of resonator A ($\omega_{\text{laser}} = \omega_{\text{A}}$), the time dependence vanishes, yielding a constant input vector $\tilde{\mathbf{S}}_{\text{in}}$ with a single nonzero entry corresponding to the $\tilde{a}_0$ mode.

In the steady state ($d\mathbf{\Phi}/dt = 0$), the linear system is solved as
\begin{equation}
    \mathbf{\Phi}_{\text{ss}} = -i \sqrt{\kappa_{\text{ex}}} \left( H_{\text{eff}} - i\frac{\kappa}{2} \mathbf{I} \right)^{-1} \tilde{\mathbf{S}}_{\text{in}}.
\end{equation}
The steady-state output field is determined by the input-output boundary condition $\tilde{\mathbf{S}}_{\text{out}} = \tilde{\mathbf{S}}_{\text{in}} - \sqrt{\kappa_{\text{ex}}} \mathbf{\Phi}_{\text{ss}}$ (through port) or $\tilde{\mathbf{S}}_{\text{out}} = \sqrt{\kappa_{\text{ex}}} \mathbf{\Phi}_{\text{ss}}$ (drop port). The optical spectrum analyzer measures the power at each frequency mode $\omega_n$, which is proportional to the mode occupation $|\tilde{a}_n|^2$. Since $|\tilde{a}_n|^2 = |a_n|^2$, the rotating frame photon number directly corresponds to the laboratory frame observable. 

\bibliography{mybib}

\end{document}